\documentclass[twocolumn,aps,pra,superscriptaddress]{revtex4}
\usepackage{amsmath,amssymb,mathrsfs,bm}
\usepackage{graphicx,dcolumn,times}
\usepackage[colorlinks,linkcolor=blue,anchorcolor=blue,citecolor=blue,filecolor=blue,menucolor=blue,runcolor=blue,urlcolor=blue,frenchlinks=blue]{hyperref}
\usepackage{amsmath}
\usepackage{graphicx}
\usepackage{amssymb}
\usepackage{dcolumn}
\usepackage{mathrsfs}
\usepackage{bm}

\begin{document}
\title{Demonstration of the quantum principle of least action with single photons}
\author{Yong-Li Wen}
\thanks{These authors contributed equally.}

\affiliation{Guangdong Provincial Key Laboratory of Quantum Engineering and Quantum Materials, School of Physics and Telecommunication Engineering, South China Normal University, Guangzhou 510006, China}
\affiliation{ Guangdong-Hong Kong Joint Laboratory of Quantum Matter, Frontier Research Institute for Physics, South China Normal University, Guangzhou 510006, China}
\affiliation{National Laboratory of Solid State Microstructures and School of Physics, Nanjing University, Nanjing 210093, China}

\author{Yunfei Wang}
\thanks{These authors contributed equally.}

\affiliation{Guangdong Provincial Key Laboratory of Quantum Engineering and Quantum Materials, School of Physics and Telecommunication Engineering, South China Normal University, Guangzhou 510006, China}
\affiliation{ Guangdong-Hong Kong Joint Laboratory of Quantum Matter, Frontier Research Institute for Physics, South China Normal University, Guangzhou 510006, China}

\author{Li-Man Tian}
\thanks{These authors contributed equally.}
\affiliation{Guangdong Provincial Key Laboratory of Quantum Engineering and Quantum Materials, School of Physics and Telecommunication Engineering, South China Normal University, Guangzhou 510006, China}
\affiliation{ Guangdong-Hong Kong Joint Laboratory of Quantum Matter, Frontier Research Institute for Physics, South China Normal University, Guangzhou 510006, China}

\author{Shanchao Zhang}

\affiliation{Guangdong Provincial Key Laboratory of Quantum Engineering and Quantum Materials, School of Physics and Telecommunication Engineering, South China Normal University, Guangzhou 510006, China}
\affiliation{ Guangdong-Hong Kong Joint Laboratory of Quantum Matter, Frontier Research Institute for Physics, South China Normal University, Guangzhou 510006, China}

\author{Jianfeng Li}

\affiliation{Guangdong Provincial Key Laboratory of Quantum Engineering and Quantum Materials, School of Physics and Telecommunication Engineering, South China Normal University, Guangzhou 510006, China}
\affiliation{ Guangdong-Hong Kong Joint Laboratory of Quantum Matter, Frontier Research Institute for Physics, South China Normal University, Guangzhou 510006, China}

\author{Jing-Song Du}

\affiliation{Guangdong Provincial Key Laboratory of Quantum Engineering and Quantum Materials, School of Physics and Telecommunication Engineering, South China Normal University, Guangzhou 510006, China}
\affiliation{ Guangdong-Hong Kong Joint Laboratory of Quantum Matter, Frontier Research Institute for Physics, South China Normal University, Guangzhou 510006, China}

\author{Hui Yan}
\email{yanhui@scnu.edu.cn}
\affiliation{Guangdong Provincial Key Laboratory of Quantum Engineering and Quantum Materials, School of Physics and Telecommunication Engineering, South China Normal University, Guangzhou 510006, China}
\affiliation{ Guangdong-Hong Kong Joint Laboratory of Quantum Matter, Frontier Research Institute for Physics, South China Normal University, Guangzhou 510006, China}

\author{Shi-Liang Zhu}
\email{slzhu@scnu.edu.cn}

\affiliation{Guangdong Provincial Key Laboratory of Quantum Engineering and Quantum Materials, School of Physics and Telecommunication Engineering, South China Normal University, Guangzhou 510006, China}
\affiliation{ Guangdong-Hong Kong Joint Laboratory of Quantum Matter, Frontier Research Institute for Physics, South China Normal University, Guangzhou 510006, China}

\begin{abstract}
The principle of least action is arguably the most fundamental principle in physics as it can be used to derive the equations of motion in various branches of physics. However, this principle  has not been experimentally demonstrated at the quantum level because the propagators for Feymann's path integrals have never been observed. The propagator is a fundamental concept and contains various significant properties of a quantum system in path integral  formulation, so its  experimental observation  is itself essential in quantum mechanics. Here we theoretically propose and experimentally observe single photons' propagators based on the method of directly measuring quantum wave-functions. Furthermore, we obtain the classical trajectories of the single photons in free space and in a harmonic trap based on the extremum of the observed propagators, thereby experimentally demonstrating the quantum principle of least action. Our work  paves the way for experimentally exploring fundamental problems of quantum theory in the formulation of path integrals.
\end{abstract}

\maketitle

The principle of least action (PLA) is a variational principle that, when applied to the action of a mechanical system, can be used to obtain the equations of motion for that system. The PLA is possibly the most  fundamental principle in physics because the fundamental laws in various branches of physics, such as classical mechanics, electrodynamics, special and general relativity, quantum mechanics, and quantum field theory, can be derived from the PLA  \cite{Rojo2018,Feynman1948,Feynman2010}.  Historically, the PLA is used in several different contexts, such as Hamilton's principle and Maupertuis' principle in classical mechanics, and Fermat's principle of least time in optics.  Intriguingly, although Einstein did not follow a least action approach in his theories of relativity, Planck formulated the dynamics of the special relativity using the PLA in 1907. More interestingly, after Hilbert learned about Einstein's initial idea of general relativity, he followed the PLA approach, guessed the ``most natural" Lagrangian in 1915, and derived the gravitational field equations before Einstein  did\cite{Rojo2018}. By extending the PLA to quantum mechanics,  Feynman discovered the path integral formulation of quantum mechanics,  which is a crucial representation of quantum mechanics and has  profoundly advanced the development of theoretical physics \cite{Feynman1948,Feynman2010}.

For non-quantum systems, an experimental demonstration of the PLA is easy since the trajectory of the system can be readily observed. However, this principle  has not been experimentally  demonstrated at the quantum level because of two obstacles.  One is conceptual obstacle, and the other is technical. Conceptually, one can determine the unique trajectory of a classical object moving from an initial position to a final position, but in quantum mechanics, according to Heisenberg's uncertainty principle, the position and the momentum cannot be simultaneously measured. Furthermore, any path connected with initial and final points is possible in quantum mechanics, so a unique trajectory like that in non-quantum physics does not exist. Feynman's path integral formulation creates a bridge between the classical Lagrangian description of the physical world and the quantum one, reintroducing to quantum mechanics the classical concept of trajectory. Technically, the PLA in Feymann's path integrals is associated with a core quantity called propagator, which has never been experimentally observed.
  This propagator is a complex amplitude with both real and imaginary components, and thus can not be measured by conventional projective measurement schemes. Recently, a method known as `direct measurement' was developed \cite{Lundeen2011} to measure the quantum wave-functions  \cite{Salvail2013,Thekkadath2016,Lundeen2012,Malik2014,Shi2015,Bolduc2016,Vallone2016,SCZhang2019}. With this method,  the real and imaginary components of the quantum wave-function can be directly read from the measuring equipment based on an interesting concept called weak value \cite{Aharonov1988,Ritchie1991,Dressel2014}. Weak value has been widely used in  precision metrology \cite{Hosten2008,Dixon2009,Hallaji2017}  and observation of
nonclassical paths \cite{Weber2014,Murch2013,Kocsis2011,YPan2020}.

In this article, we report the first experiment to  measure single photons' propagators in Feynman's path integrals and then demonstrate the PLA at the quantum level.   We theoretically develop a `direct measurement' method to observe the propagators (in the sense of directly reading them  from  the experimental apparatus) and then experimentally adopt this approach to measure the propagators' real and imaginary components. Furthermore, by analysing the extremum of the measured propagators, we experimentally demonstrate the PLA in Feynman's path integrals with single photons.

 \begin{figure*}
\includegraphics[width=12cm]{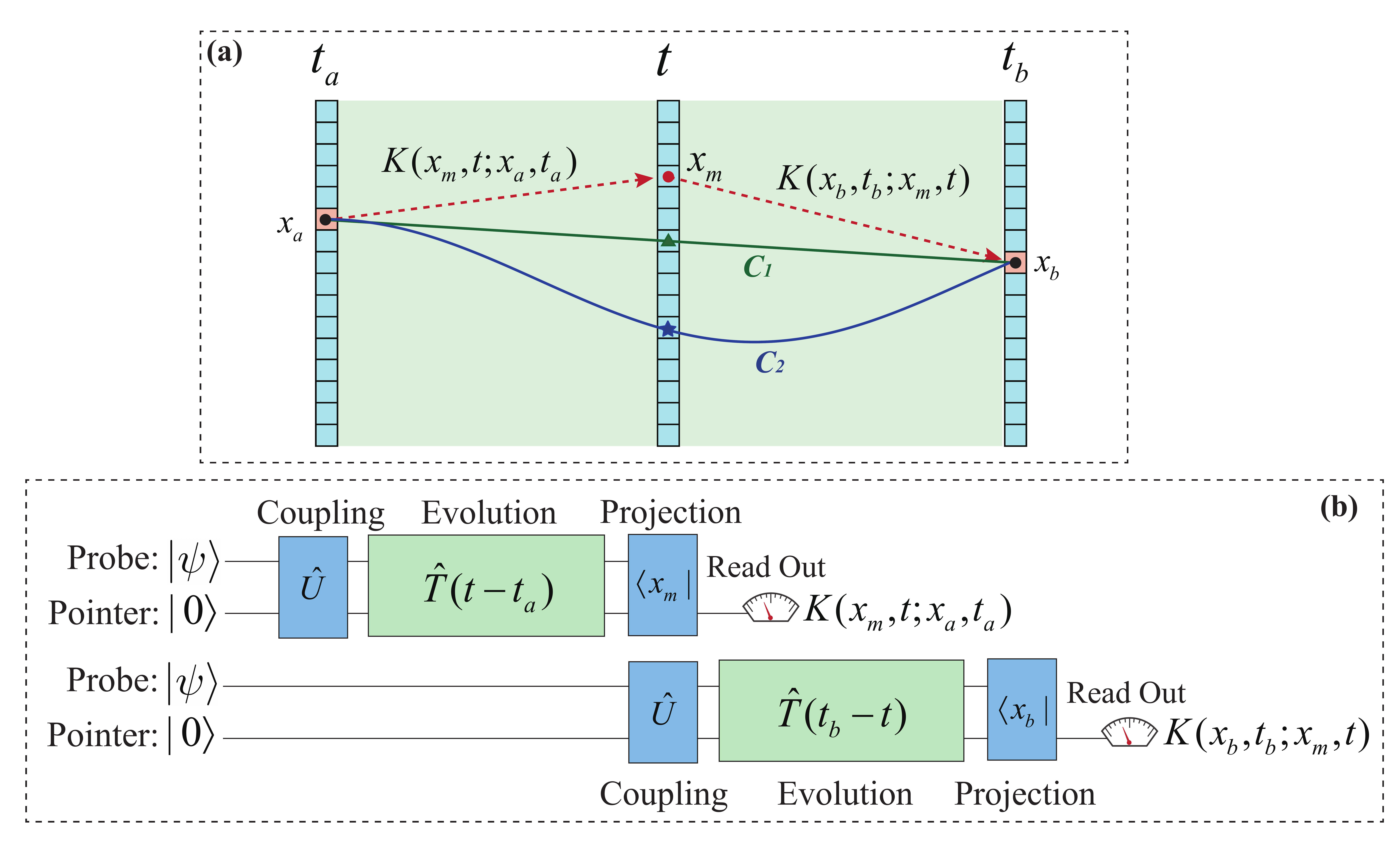}\\
\caption{\label{Theory model}\textbf{Schematic representation of path integrals and measuring method. a}, Propagators and classical trajectories. $C_1$ $(C_2)$ is the classical trajectory in free space (harmonic potential). The propagation from $(x_a,t_a)$ to $(x_b,t_b)$ is divided into two parts with intermediate points $(x,t)$. The product of propagators $K(x_b,t_b;x,t)K(x,t;x_{a},t_{a})$  is used to determine  the classical trajectory.  The classical position in free space (harmonic potential) determined by Eq.(\ref{PLA}) is shown as the green triangle (blue star). \textbf{b}, The protocol to measure the propagators. For the measurement of $K(x_m,t;x_a,t_a)$, the quantum state $|\psi\rangle$  and pointer $|0\rangle$ are prepared. Then, at  time $t_a$, a coupling operation $\hat{U}$ allows the probe system to shift the pointer at  position $x_a$. After the evolution, the system is projected to position $x_m$ at $t$, and $K(x_m,t;x_a,t_a)$ can be read out from the pointer. Similarly, $K(x_b,t_b;x_m,t)$ can be measured by performing $\hat{U}$ at $t$ and projecting the system to $x_b$ at $t_b$.}
\end{figure*}

\section{Results}
\subsection{Theoretical model}

 We use Fig.~\ref{Theory model}a to review some basic concepts of path integrals. We assume an initial wave function $|\psi (x_a,t_a)\rangle$ at a fixed position $x_a$ and  time $t_a$ and use them to  determine the wave function $|\psi (x_b,t_b)\rangle$ at another fixed position $x_b$ and  time $t_b$.  $|\psi (x_b,t_b)\rangle= K(x_{b},t_{b};x_{a},t_{a})|\psi (x_a,t_a)\rangle$ in the path integral formulation, where the propagator $K(x_{b},t_{b};x_{a},t_{a})$ describes the transition from the initial state $|\psi (x_a,t_a)\rangle$ to the final state $|\psi (x_b,t_b)\rangle$. Given an arbitrary position $x$ at the intermediate time $t$, the propagator $K(x_{b},t_{b};x_{a},t_{a})$ can be obtained by an integration,

 \begin{figure*}
\centering
\includegraphics[width=16cm]{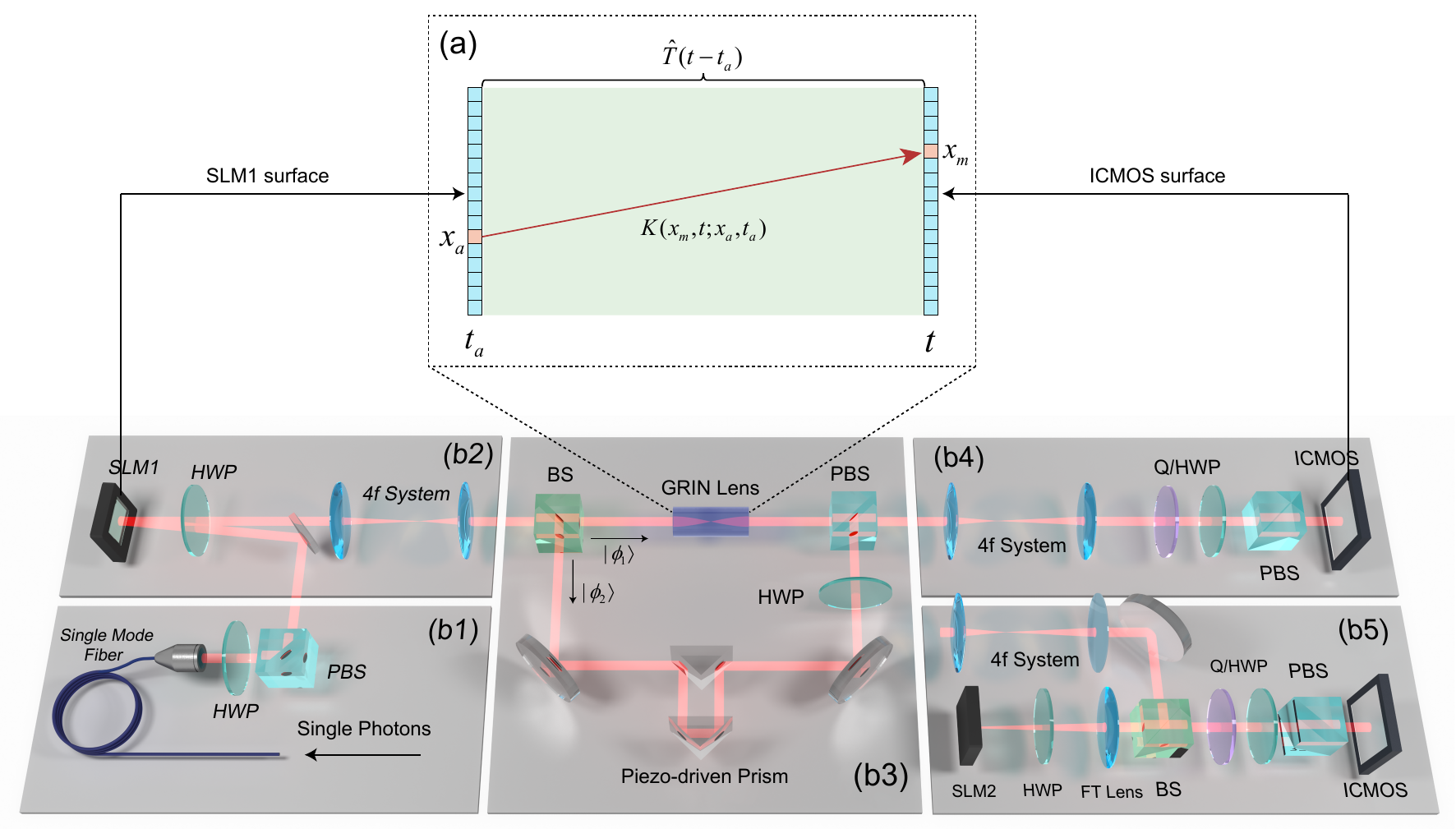}\\
\caption{\label{Experimental_setup}\textbf{Schematic representation of the experiment.} \textbf{a}, The evolution and detection region. For the measurement of $K(x_m,t;x_a,t_a)$, a spatial light modulator (SLM1) accomplishes a coupling $\hat{U}$ at  position $x_a$, while the ICMOS camera measures the photon spatial distribution at all positions $x$ for $z=ct$.
\textbf{b1-b5}, Experimental setup.
 \textbf{b1}, State initialization: single photons with a Gaussian transverse mode  emerge from a single-mode fiber and are initialized to vertical polarization state $|0\rangle$.
  \textbf{b2}, Spatial-pointer coupling:
   the combination of the HWP, SLM1 and 4f-system  realizes the $\hat{U}$ operation only at assigned positions, such as $x_{a}$.
   \textbf{b3}, State evolution: the photons are split into two branches $|\phi_1\rangle$ and $|\phi_2\rangle$ by a 50:50 beam splitter (50:50 BS). The $|\phi_1\rangle$ branch is sent into the region with the Hamiltonian $H$ corresponding to the measured propagator $K(x_m,t;x_{a},t_{a})$. A gradient refractive index (GRIN) lens produces a harmonic potential, and the absence of the GRIN lens implies free propagation.
     After the evolution, these two branches are merged by a PBS. The piezo-driven prism is used to stabilize the optical path difference between these two branches. \textbf{b4}, Post-selection of position and the readout of the pointer: a $4f$-system transits the wave front at the measured positions ($z=ct$ or $z=ct_b$) to the ICMOS  camera. Each detection of the photons on the ICMOS camera is gated (see  Supplementary Information Section 2 \cite{Supple}). Different image planes of $t$ can be measured by adjusting the longitudinal position of the ICMOS camera. The quarter-wave plate or half-wave plate (Q/HWP) and the PBS between the $4f$-system and ICMOS camera are used to read out the expectations of the pointer. \textbf{b5}, Measurement of the wave function: The combination of a FT-Lens, a HWP, and a SLM (HOLOEYE, PLUTO-2-NIR-011) accomplishes a coupling between the transverse momentum and the polarization of the photons. The ICMOS camera measures the spatial distribution of the photons of the five bases of polarization, $\{|+\rangle,|-\rangle,|R\rangle,|L\rangle,|1\rangle\}$.}
\end{figure*}

\begin{equation}\label{propagator}
K(x_{b},t_{b};x_{a},t_{a})
=\int \Pi (x,t)dx
=\mathcal{N}\mathrm{e}^{(i / \hbar) S_{cl}},
\end{equation}
where $\Pi (x,t)\equiv K(x_{b},t_{b};x,t)K(x,t;x_a,t_a)$,  $S_{cl}$ is the classical action of the path, and $\mathcal{N}$ is a normalization constant. The classical trajectory  complies with the PLA and satisfies the variational equation of $\delta S/\delta x(t)=0$ with $S$ being the action. When the ratio of  action $S$ to the Planck constant $\hbar$ increases, the phase factor $\mathrm{e}^{(i / \hbar) S}$  behaves as a strong oscillatory function and, according to a heuristic extrapolation of the stationary phase method to the path integral case, the main contribution  should come from those paths that make the phase function stationary. Therefore, one can derive a classical-like  trajectory of the system from the PLA.

The variational formulation  of the PLA is difficult to demonstrate in experiments; however, for fixed initial and final positions $x_a$ and $x_b$, the stationary of the action $S$ leads to
\begin{equation}\label{PLA}
\begin{split}
\frac{\partial}{\partial x}\Pi (x,t)=0
\end{split}
\end{equation}
for any time $t$, so the classical trajectory at the intermediate position $x$ should satisfy Eq. (\ref{PLA}), and this form of the PLA can be readily demonstrated in experiments. The analytic expressions of the propagators for a particle in free space and in a harmonic potential are provided in Methods.  Substituting these expressions in Eq. (\ref{PLA}), we  obtain that the classical trajectory is a straight line (cosine type of curve) for a particle moving from $(x_a,t_a)$ to $(x_b,t_b)$ in free space (a harmonic potential),  as shown in Fig.~\ref{Theory model}a.  Therefore, by changing the intermediate time $t$, the classical trajectory connecting two points  can be derived with Eq. (\ref{PLA}), and the PLA at the quantum level can be demonstrated  if the propagators are measured.

We now describe our approach to measure  single photons' propagators and to demonstrate the PLA. The schematic representation of the method is shown in Fig.~\ref{Theory model}. For simplicity, we assume that the photons propagate along the $z$ direction and that the initial wave function is a Gaussian wave packet so we can focus on the $x-z$ plane. As shown in Methods, the system we studied can be considered  as a one-dimensional system with $x$ as the position coordinate and $z$ as the propagation time with $t=z/c$ ($c$ is the velocity of light). We use the propagator $K(x,t;x_a,t_a)$ as an example to describe our measurement method.

  We take the spatial mode as the explored system and  the polarization of the photons as the pointer. The initial state is $|\psi\rangle|0\rangle$ where $|0\rangle$ is the pointer state. At the initial moment $t_a$, we use a coupling operation $\hat{U}=e^{-i\frac{\pi}{2}\hat{\pi}_{x_a}\hat{\sigma}_{y}}$ to couple the spatial mode of the single photons to the pointer. Here $\hat{\pi}_{x}=|x\rangle\langle x|$ and $\hat{\sigma}_{x,y,z}$ are Pauli matrixes acting on the pointer states.  This coupling only allows the photons at the position  $x_a$ shifting the pointer.
  Then, the photons enter the region with a potential $V(x)$  for the evolution ${\hat{T}}(t-t_a)=\exp(-\frac{i}{\hbar}\int_{t_a}^t Hdt')$ governed by the Hamiltonian $H=p_x^2/2m+V(x)$. In our experiment, we choose either free space or harmonic potential for this evolution. We terminate the evolution by measuring the pointer state at  position $x_m$ and time $t=z/c$. In Methods, we show that the propagator $K(x_m,t;x_{a},t_{a})$ can be obtained by
\begin{equation}\label{readout}
\begin{split}
K(x_m,t;x_{a},t_{a})
=\frac{K^{\prime\prime}(x_m,t;x_{a},t_{a})}{\sqrt{2}\psi^{*}(x_m,t_{a})\psi(x_{a},t_{a})},
\end{split}
\end{equation}
where $K^{\prime\prime}(x_m,t;x_{a},t_{a})=-\langle f|\hat{\sigma}_{x}|f\rangle+i\langle f|\hat{\sigma}_{y}|f\rangle$ with  $|f\rangle$ denoting the final state of the pointer. $\psi^{*}(x_m,t_{a})$ and $\psi(x_{a},t_{a})$ are the spatial wave functions of the photons, which can be measured by the `direct measurement' method \cite{Lundeen2011,Shi2015}. Scanning the projecting position $x_m$ allows us to measure $K(x,t;x_{a},t_{a})$ as a function of $x$. Similarly, we can derive the propagator $K(x_b,t_b;x,t)$ by scanning the coupling position $x_m$ at the moment $t$, and the projection position $x_b$ is fixed. With the measured $K(x,t;x_{a},t_{a})$ and $K(x_b,t_b;x,t)$ as a function of $x$, we use Eq.(\ref{PLA}) based on the PLA to analyse the extremum of the product  $\Pi (x,t)=K(x_b,t_b;x,t)K(x,t;x_{a},t_{a})$ and to derive classical trajectories of single photons.

\begin{figure*}
\begin{center}
\includegraphics[width=12cm]{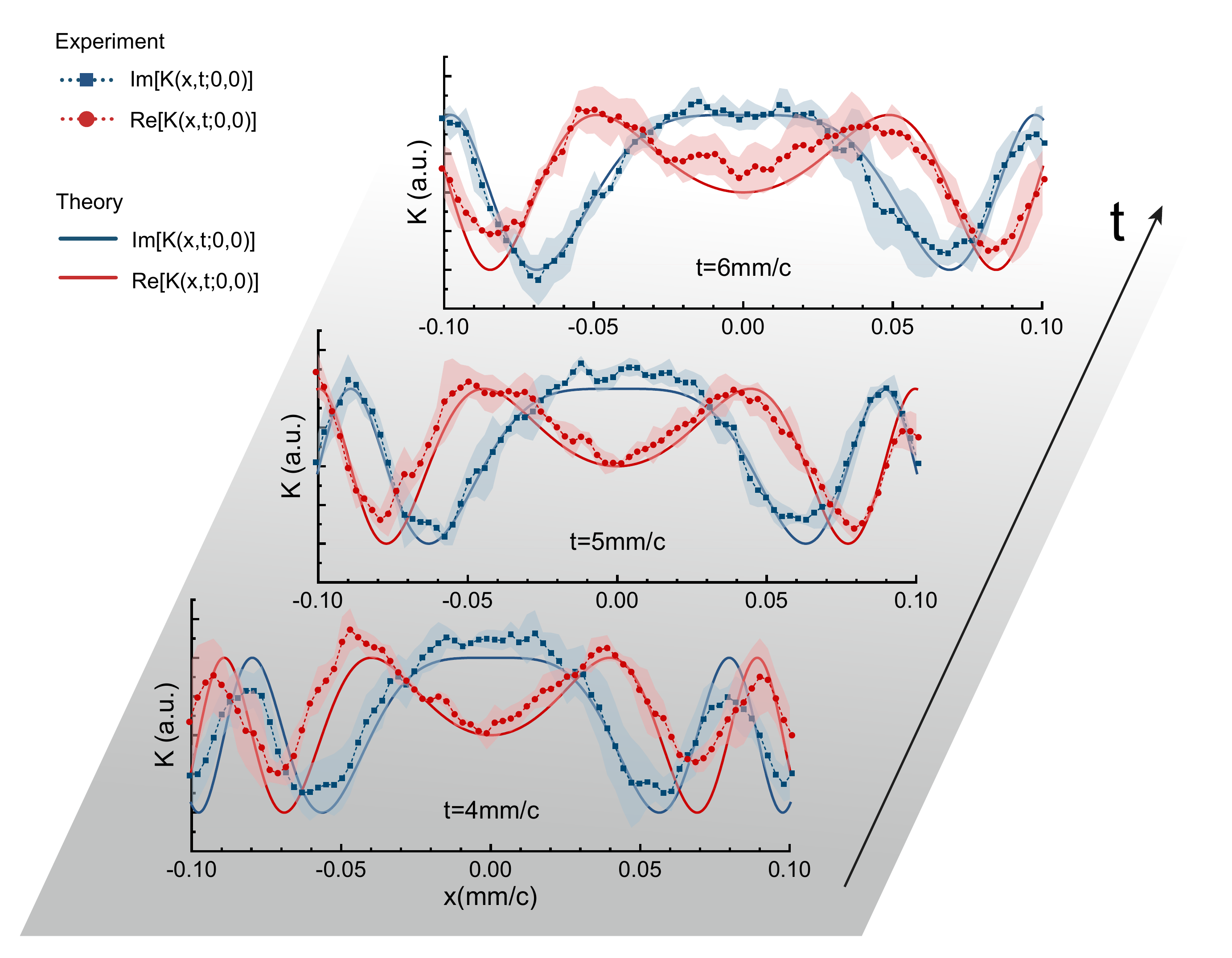}
\caption{\label{Propagator data} \textbf{Measured propagators of single photons in free space}.  The propagators $K(x,t;0,0)$  with arbitrary unit (a.u.) as a function of $x$ at $t=\{4~mm/c,5~mm/c,6~mm/c\}$.  Solid lines are  theoretical results of the real (red) and imaginary (blue) propagators. Red circles (blue squares) and the shaded error bands represent the mean value and standard deviation of three repetitive measurements.}
\end{center}
\end{figure*}

\subsection{Experimental measurement of propagators}

We perform an experiment to measure the propagators $K(x,t;x_a,t_a)$ and\\ $K(x_b,t_b;x,t)$  of single photons. The schematic of our experiment is shown in Fig.~\ref{Experimental_setup}. In our experiment, single photons  are produced through spontaneous parametric down-conversion \cite{Kwiat1995,Burnham1970,Li2021} (SPDC), with the second order correlation function $g^{(2)}_c=0.094\pm 0.011$  and a center wave length of $\lambda=795~nm$ (see Methods \ref{spdc_sec}).

We use a half-wave plate (HWP) and a spatial light modulator (SLM1; HOLOEYE, PLUTO-2-NIR-080) to couple the transverse spatial wave function to the pointer. The SLM1 rotates the polarization of photons by $\pi/2$ only at position $x_{a}$. This modulation corresponds to the unitary evolution $\hat{U}=e^{-i\frac{\pi}{2}\hat{\pi}_{x_a}\hat{\sigma}_{y}}$.  We assume that the moment right after the modulation on the SLM1 is $t_{a}$. The modulated photons then undergo a $4f$-system and are split into two paths. One of these paths passes through the region with an evolution $\hat{T}(t-t_a)$, and the other path is virtually stationary. The $4f$-system transits the spatial mode of the photons from the surface of the SLM1 to the evolution area. As shown in Fig.~\ref{Experimental_setup}b, the spatial-pointer coupling takes place at moment $t_a$. In the evolution area, the effective  potential $V(x)$ is proportional to refractive index. Thus we use a material with a quadric refractive index distribution to realize a harmonic potential for single photons and use air with uniform refractive index to achieve a free space. Another $4f$-system transits the spatial mode at $t$ to an intensified complementary metal-oxide-semiconductor (ICMOS) camera (CISS, 2DSPC; {quantum efficiency: 32\%,pixel size: 9 $\mu$m, readout noise: 4.68$e^{-}$/pix/s}). This operation realizes the projection of $x$, since the  ICMOS camera can measure the spatial distribution of photons. The positions of the pixels on the camera indicate the projecting position. Before the photons are collected by the ICMOS camera, we select the polarization with four bases: two diagonal polarization bases ($|+\rangle$ and $|-\rangle$) and two circle polarization bases ($|R\rangle$ and $|L\rangle$). The readouts from the camera on these four bases are used to calculate the expectation values in Eq. (\ref{readout}).

\begin{figure*}
\begin{center}
\includegraphics[width=12cm]{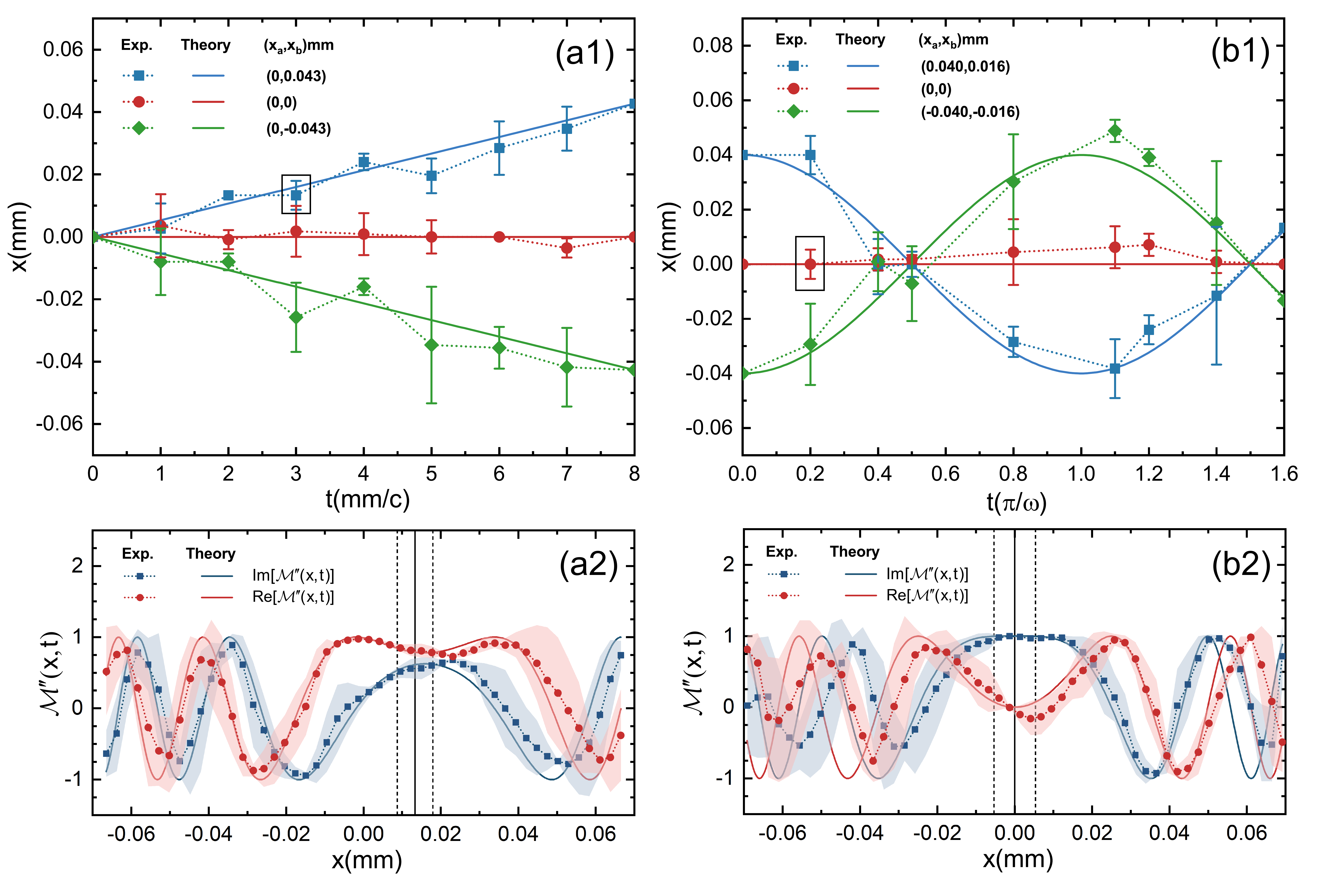}
\caption{\label{Classical path}\textbf{Classical trajectories of single photons determined by the PLA.}   \textbf{a1}, Classical trajectories in free space with fixed initial position $x_a=0$ and three final positions  $x_b=\{0.043~mm, 0, -0.043~mm\}$. \textbf{b1}, Classical trajectories  in the harmonic potential with initial position $x_a$ and final position $x_b$:  $(x_a,x_b)=(0.040~mm, 0.016~mm),(0,0), (-0.040~mm, -0.016~mm)$. Solid lines are theoretical results calculated from the analytic expressions of the propagators, and dots with error bar present mean value and standard derivation of three repetitive measurements.
\textbf{a2} and \textbf{b2},  $\mathcal{M}^{\prime\prime}(x)$ for the data in frames in \textbf{a1} and \textbf{b1}, respectively. Solid lines are  real (red) and imaginary (blue) components of the theoretical results of $\mathcal{M}^{\prime\prime}(x)$ and dots are  measured results.
The vertical solid (dashed) lines in \textbf{a2} and \textbf{b2} show the mean value (standard derivation) of classical positions determined by the measured $\mathcal{M}^{\prime\prime}(x)$. The dots with shaded error bands present the mean values and standard deviation of three repetitive measurements.}
\end{center}
\end{figure*}

We place a gradient refractive index (GRIN) lens  at the path of $|\phi_1\rangle$, where the incident surface of the GRIN lens is at the image plane of the $4f$-system (see Methods). Therefore, the GRIN lens carries out the evolution $\hat{T}_{1}(t_{b}-t_{a})$, where $t_{a}$ ($t_b$) is the incident (exit) surface. For the free propagation case, the GRIN lens is removed. The outgoing photons from this evolution then pass through a polarizing beam splitter (PBS). This PBS projects $|\phi_1\rangle$ to the pointer state $|1\rangle$ by  selecting the transmitted photons. For the other branch $|\phi_2\rangle$ we add an HWP before the PBS, and the reflected photons are projected to the state $|-\rangle$. The PBS merges these two branches  to one. The second $4f$-system after the PBS transits the state from the $t_{b}$ plane to the ICMOS camera. The camera is used to measure the spatial distribution of the photons. The intensity of the pixel at position $x_{b}$ on the ICMOS sensor represents the readout of post-selection  $|x_{b}\rangle$.  To read out the expectation value of $\langle f|\hat{\sigma}_y|f\rangle$ and $\langle f|\hat{\sigma}_x|f\rangle$, We set a quarter-wave plate or half-wave plate (QWP/HWP) and a PBS to project the photons to four polarization bases: $|+\rangle=1/\sqrt{2}(|0\rangle+|1\rangle)$, $|+\rangle=1/\sqrt{2}(|0\rangle-|1\rangle)$, $|R\rangle=1/\sqrt{2}(|0\rangle+i|1\rangle)$, and $|L\rangle=1/\sqrt{2}(|0\rangle-i|1\rangle)$. The intensities of the pixels on these four bases are $\{P_{+},P_{-},P_{R},P_{L}\}$. The difference between  $P_{+}$ and $P_{-}$ gives the expectation of $\langle f|\hat{\sigma}_x|f\rangle=P_{+}-P_{-}$, whereas the difference between $P_{R}$ and $P_{L}$ gives  the expectation of $\langle f|\hat{\sigma}_y|f\rangle=P_{R}-P_{L}$.

To measure the wave functions $\psi(x_a,t_{a})$ and $\psi(x_m,t_{a})$, we perform an interaction $\hat{U}_p=e^{-i\frac{\pi}{2}\hat{\pi}_{p_0} \hat{\sigma}_{y}}$ to couple the momentum of the photons to the pointer and then use the ICMOS camera to project the system to the position state $\langle x_j|$ (see Fig.~\ref{Experimental_setup}b5 and Supplementary Information section 1). Similarly, using a QWP/HWP and a PBS to read out the expectation values of $\hat{\sigma}_y$, $\hat{\sigma}_x$ and $\hat{P}_1=|1\rangle\langle 1|$, we have
\begin{equation}
\psi(x_{j},t_a)=-\frac{\frac{1}{2}[\langle f_{j}|\hat{\sigma}_{x}|f_{j}\rangle+i\langle f_{j}|\hat{\sigma}_{y}| f_{j}\rangle]-\langle f_{j}|\hat{P}_{1}| f_{j}\rangle}{\Phi^{*}(0,t_a)},
\end{equation}
where $j=\{m,a\}$ is the index of the measured position (see Methods \ref{wave_func_meas}).

 The propagators of single photons in free space are shown in Fig.~\ref{Propagator data}. We choose $x_a=0$ as a fixed initial point and then measure the propagators $K(x,t;x_a,t_a)$ as a  function of the final position $x$. By adjusting the ICMOS camera on the longitudinal axis, we can detect $K(x,t;x_a,t_a)$ at different evolution time $t$. The experimental results of $K(x,t;x_a,t_a)$ as a function of $x$ and $t$ are shown in Fig.~\ref{Propagator data} at $t=\{4~mm/c, 5~mm/c, 6~mm/c\}$ with the measuring step $\delta x\approx 2.67\mu m$. The experimental data agree well with the theoretical results.

\subsection{Demonstration of the PLA}

In Methods, we show that the classical trajectories based on the PLA can be determined by
\begin{equation}
\label{PLA_M}
\frac{\partial}{\partial x} Re[\mathcal{M}^{\prime\prime}(x,t)]=0,\ \ \ \frac{\partial}{\partial x} Im[\mathcal{M}^{\prime\prime}(x,t)]=0,
\end{equation}
where $\mathcal{M}^{\prime\prime} (x,t)= \Pi^{\prime\prime} (x,t)/|\Pi^{\prime\prime} (x,t)|$ with $\Pi^{\prime\prime} (x,t)=K^{\prime\prime}\left(x_{b}, t_{b}; x, t\right)K^{\prime\prime}\left(x, t; x_{a}, t_{a}\right)$. Using the same method described in the previous section, we measure both $K^{\prime\prime}\left(x, t; x_{a}, t_{a}\right)$ and $K^{\prime\prime}\left(x_b, t_b; x, t\right)$, and the classical trajectory $x_{cl}(t)$ can be obtained by Eq.(\ref{PLA_M}) with the measured data.

 In the experiments, we select an initial position $x_a$ by choosing the transverse position of the slits on the SLM1. Then $K^{\prime\prime}(x,t;x_a,t_a)$ as a function of $x$ can be measured by the ICMOS camera. In $K^{\prime\prime}(x_b,t_b;x,t)$, $x$ is the variable initial position and $x_b$ is the fixed final position. We scan the slits on the SLM1 to accomplish the change of $x$. $K^{\prime\prime}(x_b,t_b;x,t)$ can be measured by reading out the signal of the final position $x_b$ on the ICMOS camera. With these data, we can obtain $x_{cl}(t)$ by searching the  position where $\mathcal{M}^{\prime\prime}(x,t)$ has its extremum. The GRIN lens is used to simulate the harmonic potential, which can be effectively expressed as $V(x)=\frac{1}{2}m\omega^2 x^2$ (see Methods \ref{grin_met}). The GRIN lens is removed when propagators of the free particle are measured.

  The classical trajectories determined by the measured propagators are plotted in Fig.~\ref{Classical path}, where they are a straight line (cosine type of curve) for a particle moving from $(x_a,t_a)$ to $(x_b,t_b)$ in free space (a harmonic potential), as expected by the analytical expressions in Eqs. (\ref{Kf},\ref{Kh}). The propagators $\mathcal{M}^{\prime\prime}$ for those data in dashed frames in Figs.~\ref{Classical path}{a1} and ~\ref{Classical path}{b1} are plotted in Figs.~\ref{Classical path}{a2} and~\ref{Classical path}{b2}, respectively. The classical paths derived from the measured propagators agree well with  theoretical results, while the deviation of $\mathcal{M}^{\prime\prime}$ from the ideal result is small in the region near the classical trajectory, but large in the region far away from the classical trajectory. So the classical trajectories determined by the measured propagators are very robust.  This is also true  for various perturbations, and in Methods, we use the deviation of time as an example to show this phenomenon. This phenomenon occurs because the action $ S$ is stationary for a classical trajectory but would behave as a strong oscillation function for non-classical trajectories. Our experiment demonstrates this essential idea of the path integral theory \cite{Feynman1948,Feynman2010}.

\section{Discussion}

In summary, we have reported the first experiment to measure  single photons' propagators and to demonstrate the quantum PLA.Feynman's path integral has been recognized as a crucial theory in the development of modern quantum mechanics, particularly in quantum field theory and quantum statistical physics. The propagator, which lies at the heart of this theory, contains significant properties (such as wave function, evolution, action,topological invariants and partition function) of a quantum system. Our method of measuring the propagators provides a new perspective to study quantum systems in the path integral formulation. As our experiment shows, the classical trajectories can be obtained by measuring the propagators, which allows us to further explore the crossover between classical physics and quantum physics,  one of the  frontiers in current physics research. Moreover, the measurement of the propagators opens opportunities for experimentally exploring  physics phenomena in quantum field theory and quantum statistical physics that previously could not be experimentally observed.

\section{Methods}
\subsection{The Hamiltonian and  expressions of the propagators}

The Hamiltonian of single photons can be written as
$H=c|\vec{p}|=c\sqrt{\hat{p}_{x}^{2}+\hat{p}_{y}^{2}+\hat{p}_{z}^{2}},$
where $x$ and $y$ are two transverse coordinates, and $z$ is the propagation direction.
The momentum $p_{z}$ is related to the wave length of the light, $p_{z}=\hbar k_{z}={2\pi\hbar}/{\lambda}$.
In our experiment, the initial wave function is a Gaussian wave packet in $x-y$ plane, $|\psi({x,y})|^2=\mathcal{N}_G \exp [-({x^{2}}+{y^{2}})/({a_{x}^{2}}+{a_{y}^{2}})]$ with $\mathcal{N}_G$ as a normalization constant.
The uncertainty of the momentum $p_{x}$ ($p_{y}$) is $\hbar/a_{x}$ ($\hbar/a_{y}$). The wave length of the light we used is $\lambda=795~nm$ and $a_{x}=a_{y}\approx 0.4~mm$, so we have $p_{z} \gg p_{x}, p_{y}$. Under this condition, the Hamiltonian can be rewritten as an approximation
\begin{equation}
H\approx \frac{c\hat{p}_{x}^{2}}{2p_{z}}+\frac{c\hat{p}_{y}^{2}}{2p_{z}}+cp_{z}.
\end{equation}
 For the sake of simplicity and without loss of generality, we focus only on the $x-z$ plane. In addition, because we consider $p_{z}$ a near constant,  $z$ is proportional to the propagation time $t=z/c$. Then, the evolution of the spatial wave function $\psi(x,t)$ is described by a Schr\"{o}dinger-like equation
\begin{equation}
\label{Schrodinger-Like}
i\hbar\frac{\partial}{\partial t}\psi(x,t)=\frac{c\hat{p}_{x}^{2}}{2p_{z}}\psi(x,t).
\end{equation}
This result shows that the photon acts like a non-relativistic particle with  effective mass $m=\frac{p_{z}}{c}=\frac{2\pi\hbar}{\lambda c}$ in transverse dimensions when the propagation momentum is much greater than the uncertainty of the transverse momentum \cite{YLWen2021}.

This Schr\"{o}dinger-like equation (\ref{Schrodinger-Like}) can also be derived from the paraxial Helmholtz equation. Consider an optical wave propagates along $z$ axis,
\begin{equation}
u(x, y, z)=\psi(x, y, z) e^{i k z},
\end{equation}
where $k$ is the wave number of the propagation. The Helmholtz equation can be written as
\begin{equation}
\nabla^{2} u(x, y, z)+k^{2} u(x, y, z)=0.
\end{equation}
Assuming that $\psi(x, y, z)$ is a slowly varying function of $z$ comparing with $k$, the Helmholtz equation can be written as a paraxial approximation form,
\begin{equation}
\frac{\partial^{2}}{\partial x^{2}} \psi(x, y, z)+\frac{\partial^{2}}{\partial y^{2}} \psi(x, y, z)+2 i k \frac{\partial}{\partial z} \psi(x, y, z)=0.
\end{equation}
In quantum mechanics, the partial differentials can be transformed to momentum operators,
\begin{equation}\label{px_py}
\hat{p}_{x}=-i \hbar \frac{\partial}{\partial x},\ \ \hat{p}_{y}=-i \hbar \frac{\partial}{\partial y}.
\end{equation}
Since the wave is propagating along $z$ direction, we can consider the momentum in this direction as a constant,
\begin{equation}\label{pz}
p_{z}=\hbar k.
\end{equation}
Under this condition, the partial differential of $z$ can be also rewritten as
\begin{equation}\label{partialz}
\frac{\partial}{\partial z}=\frac{1}{c} \frac{\partial}{\partial t}.
\end{equation}
By using Eqs. (\ref{px_py})-(\ref{partialz}), and focusing only on the $x-z$ plane, the paraxial Helmholtz equation in quantum mechanics can be rewritten as a Schr\"{o}dinger-like equation
\begin{equation}
i \hbar \frac{\partial}{\partial t} \psi(x, t)=\frac{c \hat{p}_{x}^{2}}{2 p_{z}} \psi(x, t).
\end{equation}

If we choose refractive index as a function of $x$, then the photons will have an effective potential $V(x)$ and act as a non-relativistic particle with a Hamiltonian $H=p_x^2/2m+V(x)$. Therefore,
the propagator of single photons in free space for two fixed points $(x_a,t_a)$ and $(x_b,t_b)$ is given by
\begin{equation}
\label{Kf}
 K_{f}(x_{b},t_{b}; x_{a},t_{a} )=\sqrt{\frac{m}{2 \pi i \hbar t_{ba}}} \exp\left({\frac{i m(x_{b}-x_{a})^{2}}{2\hbar t_{ba}}}\right),
 \end{equation}
where $t_{ba}=t_b-t_a$. The propagator in a harmonic potential $V(x)=m\omega^{2} x^2/2$ is given by  \cite{Feynman2010},
\begin{equation}
\label{Kh}
\begin{split}
&\quad K_{h}(x_{b},t_{b};x_{a},t_{a})
\\&=\mathcal{N}_h\exp\left\{ \frac{i m\omega}{2\hbar\sin \omega t_{ba}}\left[\left(x_{a}^{2}+x_{b}^{2}\right) \cos \omega t_{ba}-2 x_{a} x_{b}\right]\right\}
\end{split}
\end{equation}
where $\mathcal{N}_h=\sqrt{\frac{m\omega}{2\pi i\hbar\sin\omega t_{ba}}}$.
\subsection{Measurement of the propagators}

We take the propagator  $K(x_{m},t; x_{a},t_{a} )$ with $x_m$ being one of the points at time $t$ as an example to describe our scheme of measuring propagators.
We first divide the transverse position into $d$ slits, so a wave function of transverse position at time $t$ can be written as $|\psi(t)\rangle=\sum^{d}_{j=1}\psi(x_j,t)|x_j\rangle$ in basis  $|x\rangle$.
We choose a 2-dimension qubit space with eigenstates $|0\rangle$ and $|1\rangle$  as the pointer. Our initial state  $|\phi_{i}\rangle$ at $t_{a}$  is prepared as
\begin{equation}
|\phi_{i}\rangle=|\psi\rangle |0\rangle .
\end{equation}
At this moment, we perform an interaction \cite{vonNeumann1955} $H_{I}= \hat{\pi}_{x_a}\hat{\sigma}_{y}$ with $\hat{\pi}_{x_a}=|x_a\rangle\langle x_a|$ on the system, and we have
\begin{equation}
U(\theta)|\phi_{i}\rangle=e^{-i\theta \hat{\pi}_{x_a} \hat{\sigma}_{y}}|\psi\rangle|0\rangle,
\end{equation}
where $\theta$ is an  angle related to the interaction strength.
It also reflects whether a measurement is strong or weak. In our case, we choose a strong interaction with  $\theta=\pi/2$. After this, the system state is given by
\begin{equation}
\label{phi_a}
|\phi_{a}\rangle=[|\psi\rangle-\psi(x_{a},t_{a})|x_{a}\rangle]|0\rangle-\psi(x_{a},t_{a})|x_{a}\rangle|1\rangle.
\end{equation}
Then the system evolves with a Hamiltonian $H=p_x^2/2m+V(x,t)$ till time $t$, and we have an evolution operator  $\hat{T}(t-t_{a})=\exp{[-(i/\hbar)\int_{t_{a}}^{t} Hdt']}$. The state becomes
\begin{equation}
\begin{split}
\label{phi_t}
|\phi_{t}\rangle=&[\hat{T}(t-t_{a})|\psi\rangle-\psi(x_{a},t_{a})\hat{T}(t-t_{a})|x_{a}\rangle]|0\rangle
\\&-\psi(x_{a},t_{a})\hat{T}(t-t_{a})|x_{a}\rangle|1\rangle,
\end{split}
\end{equation}
 At time $t$, we perform a post-selection by projecting the system to a position state $\langle x_{m}|$. The final state of the pointer $|f\rangle=\langle x_m|\phi_{t}\rangle$  can be written as
\begin{equation}
\begin{split}
|f\rangle=& [\langle x_{m}|\hat{T}(t-t_{a})|\psi\rangle
\\&-\psi(x_{a},t_{a})\langle x_{m}|\hat{T}(t-t_{a})|x_{a}\rangle]|0\rangle
\\& -\psi(x_{a},t_{a})\langle x_{m}|\hat{T}(t-t_{a})|x_{a}\rangle|1\rangle.
\end{split}
\end{equation}
Since the propagator  $K(x_{m},t;x_{a},t_{a})=\langle x_{m}|\hat{T}(t-t_{a})|x_{a}\rangle$, it can be rewritten as
\begin{equation}
\begin{split}
|f\rangle=& [\psi(x_{m},t)-\psi(x_{a},t_{a})K(x_{m},t;x_{a},t_{a})]|0\rangle
\\& -\psi(x_{a},t_{a})K(x_{m},t;x_{a},t_{a}) |1\rangle.
\end{split}
\end{equation} Here, we notice that the information of the propagator $K(x_{b},t_{b};x_{a},t_{a})$ is contained in the pointer. The method of extracting the propagator is similar to that of extracting wave function in the `direct measurement' scheme developed in Ref.\cite{Lundeen2011}. If we measure the expectation value of the operators $\hat{\sigma}_{x}$, $\hat{\sigma}_{y}$ and $\hat{P}_{1}=|1\rangle\langle1|$, we have
\begin{equation}
\begin{split}
\langle f|\hat{\sigma}^{-}|f\rangle
&=-\psi^{*}(x_{m},t)\psi(x_{a},t_{a})K(x_{m},t;x_{a},t_{a})\\
&+|\psi(x_{a},t_{a})K(x_{m},t;x_{a},t_{a})|^{2}, \\
\langle f|\hat{P}_{1}|f\rangle&=|\psi(x_{a},t_{a})K(x_{m},t;x_{a},t_{a})|^{2},
\end{split}
\end{equation}
where $ \hat{\sigma}^{-}=\frac{1}{2}(\hat{\sigma}_{x}-i\hat{\sigma}_{y})$. Finally, the propagator can be obtained as
\begin{equation}
\label{Kxx}
K(x_{m},t;x_{a},t_{a})=\frac{K'(x_m,t;,x_a,t_a)}
{\psi^{*}(x_{m},t)\psi(x_{a},t_{a})},
\end{equation}
where $K'(x_m,t;,x_a,t_a)=\frac{1}{2}[-\langle f|\hat{\sigma}_{x}|f\rangle+i\langle f|\hat{\sigma}_{y}| f\rangle]+\langle f|\hat{P}_{1}| f\rangle$. The wave functions $\psi^{*}(x_{m},t)$ and $\psi(x_{a},t_{a})$ can be measured using the `direct measurement' method described  in previous works  \cite{Lundeen2011,Shi2015,Vallone2016}, also in Section \ref{wave_func_meas}.

However, it is difficult to extract the propagator using Eq. (\ref{Kxx}) when $\psi(x_{m},t) \approx 0$.
 To solve this problem, we propose an optimized scheme. We separate the system into two branches at $t_{a}$. These two branches evolve differently and can be written as
\begin{equation}
\begin{split}
|\phi_{1}\rangle=&[\hat{T}_{1}(t-t_{a})|\psi\rangle-\psi(x_{a},t_{a})\hat{T}_{1}(t-t_{a})|x_{a}\rangle]|0\rangle
\\&-\psi(x_{a},t_{a})\hat{T}_{1}(t-t_{a})|x_{a}\rangle|1\rangle,
\\|\phi_{2}\rangle=&[\hat{T}_{2}(t-t_{a})|\psi\rangle-\psi(x_{a},t_{a})\hat{T}_{2}(t-t_{a})|x_{a}\rangle]|0\rangle
\\&-\psi(x_{a},t_{a})\hat{T}_{2}(t-t_{a})|x_{a}\rangle|1\rangle.
\end{split}
\end{equation}
 We choose $\hat{T}_{1}(t-t_{a})=\emph{e}^{-\frac{i}{\hbar}\int_{t_{a}}^{t}[\frac{p^{2}}{2m}+V(x,t')]dt'}$ as the evolution  related to the measured propagator. In other words, $|\phi_{1}\rangle$ is the probe branch, which will obtain the propagator $K(x_{m},t;x_{a},t_{a})=\langle x_{m}|\hat{T}_{1}(t-t_{a})|x_{a}\rangle$.  $|\phi_{2}\rangle$ is a reference branch, which we set  to be a purely free evolution $\hat{T}_{2}(t-t_{a})=\emph{e}^{-\frac{i}{\hbar}\int_{t_{a}}^{t}\frac{p^{2}}{2m}}dt'$ . We rotate $|\phi_{2}\rangle$ with a unitary operator, $\emph{e}^{i\frac{\pi}{4}\hat{\sigma}_{y}}|\phi_{2}\rangle$. We then perform the projection operator of state $|1\rangle$ on $|\phi_{1}\rangle$, and perform the projection operator of state $|0\rangle$ on $\emph{e}^{i\frac{\pi}{4}\hat{\sigma}_{y}}|\phi_{2}\rangle$. The results can be written as
\begin{equation}
\begin{split}
|\phi'_{1}\rangle &=|1\rangle\langle1|\phi_{1}\rangle \\&= -\psi(x_{a},t_{a})\hat{T}_{1}(t-t_{a})|x_{a}\rangle|1\rangle
\\|\phi'_{2}\rangle &=|0\rangle\langle 0| \emph{e}^{i\frac{\pi}{4}\hat{\sigma}_{y}}|\phi_{2}\rangle\\&=\frac{1}{\sqrt{2}}\hat{T}_{2}(t-t_{a})|\psi(x_{a},t_a)\rangle|0\rangle.
\end{split}
\end{equation}
After that, we merge the two branches
\begin{equation}
\begin{split}
|\phi'\rangle =&|\phi'_{1}\rangle+ |\phi'_{2}\rangle
\\=&\frac{1}{\sqrt{2}}\hat{T}_{2}(t-t_{a})|\psi\rangle|0\rangle\\&-\psi(x_{a},t_{a})\hat{T}_{1}(t-t_{a})|x_{a}\rangle|1\rangle.
\end{split}
\end{equation}
Then we project $|\phi'\rangle$ on the post-selected position state $\langle x_{m}|$, and the result is given by
\begin{equation}
\begin{split}
|f'\rangle=&\langle x_{m}|\phi'\rangle
\\=&\frac{1}{\sqrt{2}}\langle x_{m}|\hat{T}_{2}(t-t_{a})|\psi\rangle|0\rangle\\&-\psi(x_{a},t_{a})K(x_{m},t;x_{a},t_{a})|1\rangle .
\end{split}
\end{equation}
If the free evolution $\hat{T}_{2}(t-t_{a})|\psi\rangle$ is virtually stationary, we can obtain an approximate result that $\langle x_{m}|\hat{T}_{2}(t-t_{a})|\psi\rangle\approx \psi(x_{m},t_{a})$. We have checked this approximation and found it is well satisfied in our experiments. Finally, the propagator $K(x_{m},t;x_{a},t_{a})$ can be obtained as
\begin{equation}
\label{Kxx2}
K(x_{m},t;x_{a},t_{a})
=-\frac{K^{\prime\prime}(x_{m},t;x_{a},t_{a})}{\sqrt{2}\psi^{*}(x_{m},t_{a})\psi(x_{a},t_{a})}.
\end{equation}
where $K^{\prime\prime}(x_{m},t;x_{a},t_{a})=-\langle f'|\hat{\sigma}_{x}|f'\rangle+i\langle f'|\hat{\sigma}_{y}|f'\rangle$. In this result, instead of measuring both $\psi(x_{m},t)$ and $\psi(x_{a},t_{a})$,  we just need to measure the wave functions at $t_{a}$, which do not vanish in our experimental conditions. We adopt this method in our experiments.

\subsection{Measurement of the wave functions}\label{wave_func_meas}

\begin{figure}
\begin{center}
\includegraphics[width=8cm]{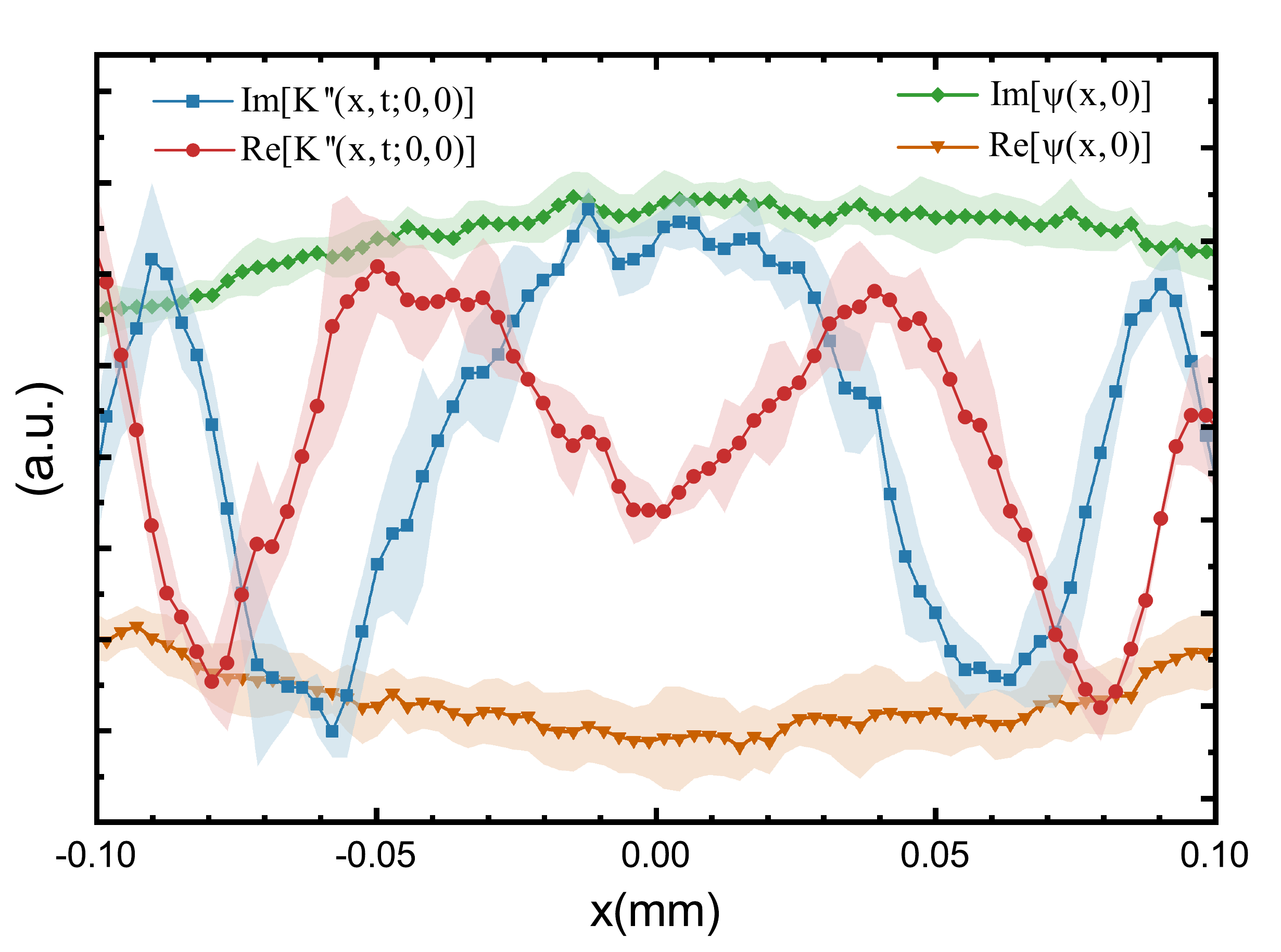}
\caption{\label{wave_function}The measured wave function $\psi(x,0)$ and $K^{\prime\prime}(x,t=5mm/c;0,0)$ with arbitrary unit (a.u.). The dots with shaded error bands represent the mean values and standard deviation of three measured results.}
\end{center}
\end{figure}

\begin{figure*}
\begin{center}
\includegraphics[width=14.0cm]{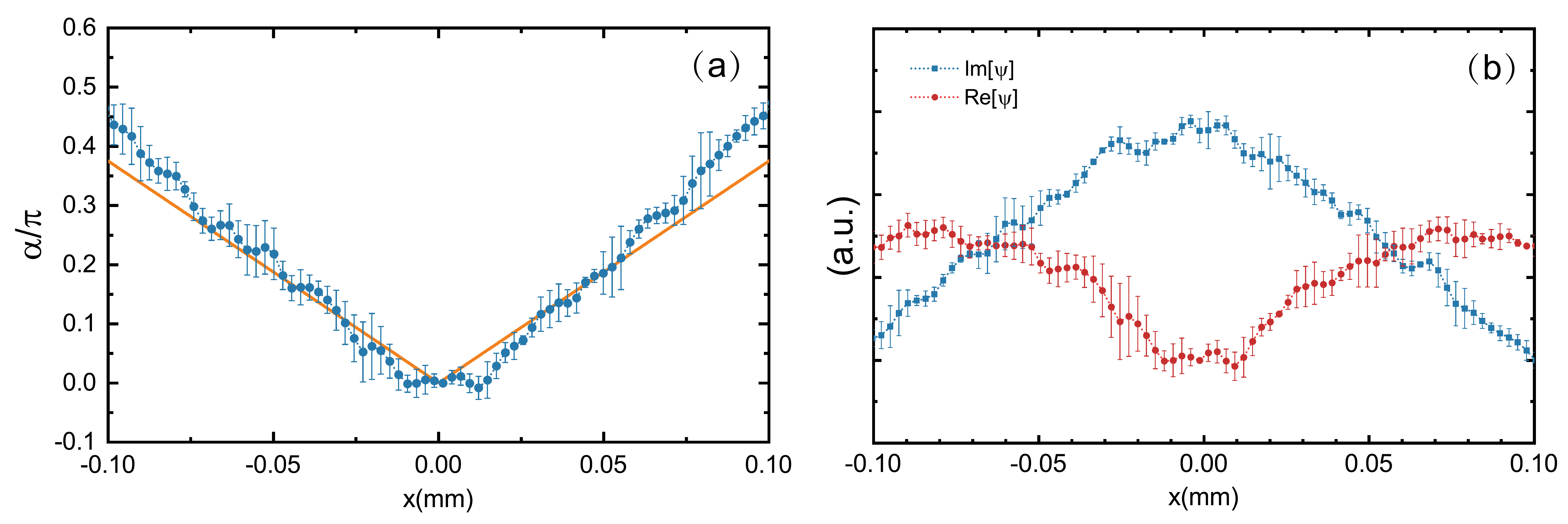}
\caption{\label{phase_mod}
Quantum wave functions with a V-shaped phase modulation. \textbf{a,}Blue circles (mean values) with error bars (standard deviation) show the measured phase profile $\alpha(x)=\arctan\{Re[\psi(x,0)]/Im[\psi(x,0)]\}$, and solid orange line shows the modulation function of the phase profile. \textbf{b}, Real and imaginary components of $\psi(x,0)$.The dots with error bar represent the mean values and standard deviation of three measured results.}
\end{center}
\end{figure*}

We adopt the  method in Ref.\cite{Shi2015} to measure the wave functions.  At the moment $t_a$, we perform an interaction $\hat{H}_p=\hat{\pi}_{p_0}{\sigma_y}$ with $\hat{\pi}_{p_0}=|p_0\rangle\langle p_0|$, where $|p_0\rangle$ is the state of zero transverse momentum. Then we have,
\begin{equation}
\hat{U}_p(\theta)|\phi_{i}\rangle=e^{-i\theta \hat{\pi}_{p_0} {\sigma}_{y}}|\psi\rangle|0\rangle.
\end{equation}
In our case, the interacting strength is set to be $\theta=\pi/2$, the state of the system is
\begin{equation}
|\phi_{p}\rangle=[|\psi\rangle-\Phi(0,t_a)|p_0\rangle]|0\rangle-\Phi(0,t_a)|p_0\rangle|1\rangle.
\end{equation}
where $\Phi(0,t_a)$ is the momentum wave function at $p_0$. Then we project the $|\phi_{p}\rangle$ onto post-selected position state $| x_j\rangle$ at moment $t_a$, the final pointer state can be written as,
\begin{equation}
\begin{split}
|f_{p}\rangle=&[\psi(x_j,t_a)-\Phi(0,t_a)e^{-ip_0 x_j/\hbar}]|0\rangle\\
&-\Phi(0,t_a)e^{-ip_0 x_j/\hbar}|1\rangle,
\end{split}
\end{equation}

where $j=\{m,a\}$ is the index of the measured position. For $p_0=0$, the phase term $e^{-ip_{0} x_{j}/\hbar}=1$ and the momentum wave function $\Phi(0,t_{a})$ is constant. The wave function $\psi(x_j,t_a)$ is obtained as,
\begin{equation}
\psi(x_{j},t_a)=-\frac{\frac{1}{2}[\langle f_{p}|\hat{\sigma}_{x}|f_{p}\rangle+i\langle f_{p}|\hat{\sigma}_{y}| f_{p}\rangle]-\langle f_{p}|\hat{P}_{1}| f_{p}\rangle}{\Phi^{*}(0,t_a)}.
\end{equation}
Similarly, if the post-selection of $|x_j\rangle $ is performed at $t$, the wave function $\psi(x_{j},t)$ can be measured by the same method.

Figure \ref{wave_function} shows the measured $K^{\prime\prime}(x,t; 0,0)$  for $t=5mm/c$ and  wave function $\psi(x,0)$ as a function of $x$.
The propagators $K(x,t;0,0)$ with $t=5mm/c$ shown in Fig.3 in the main text are obtained with the data here.

To check that a correct quantum wave function can be measured in our experiment, we perform a V-shaped phase modulation on the initial wave function, and our measured results plotted in Fig.\ref{phase_mod} show the correct phase modulation.

\subsection{The SPDC single photon source}\label{spdc_sec}
\begin{figure*}[ptb]
\begin{center}
\includegraphics[width=12cm]{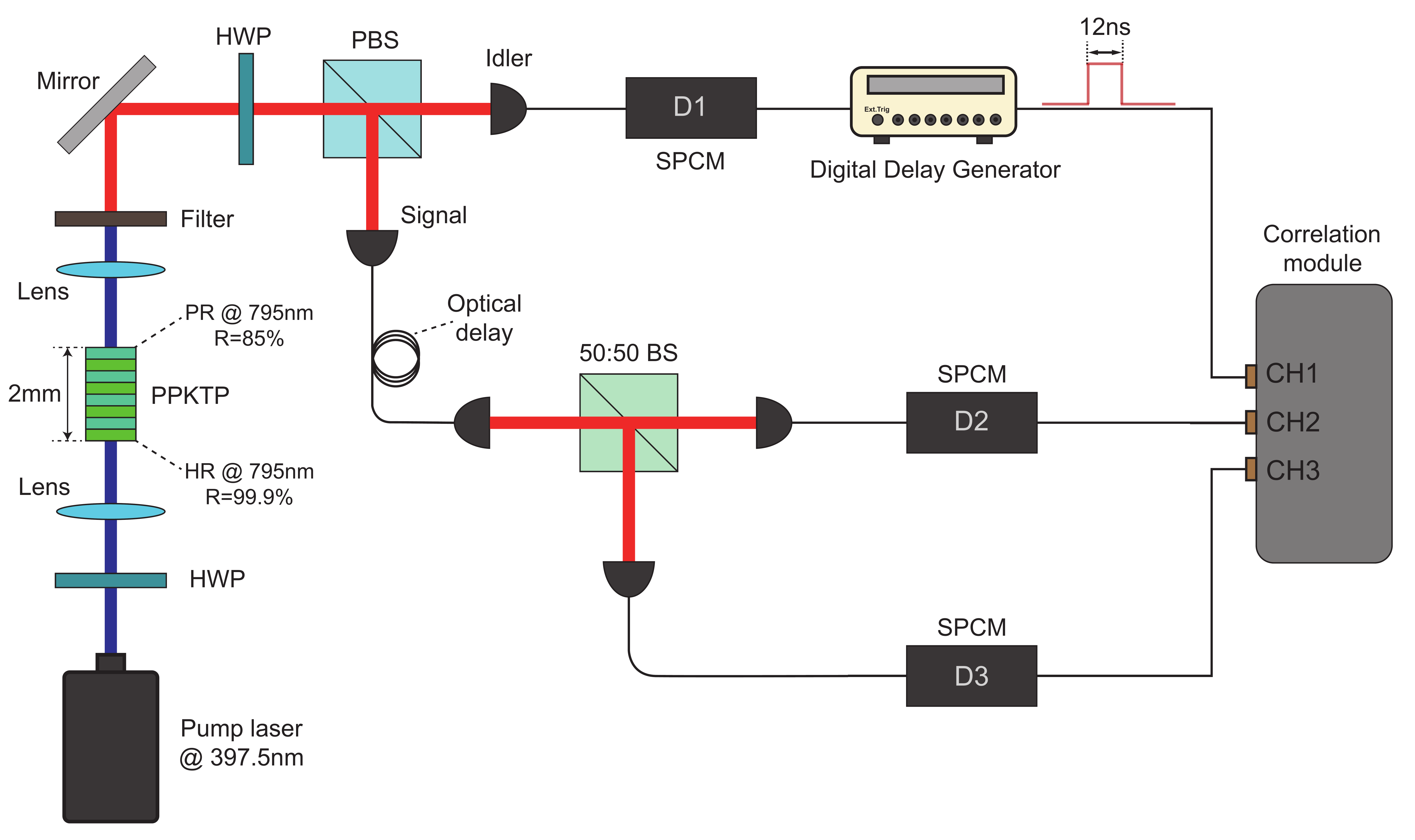}
\caption{\label{spdc} The experimental schematic of the SPDC single photon source. }
\end{center}
\end{figure*}

\begin{figure}[ptb]
\begin{center}
\includegraphics[width=6cm]{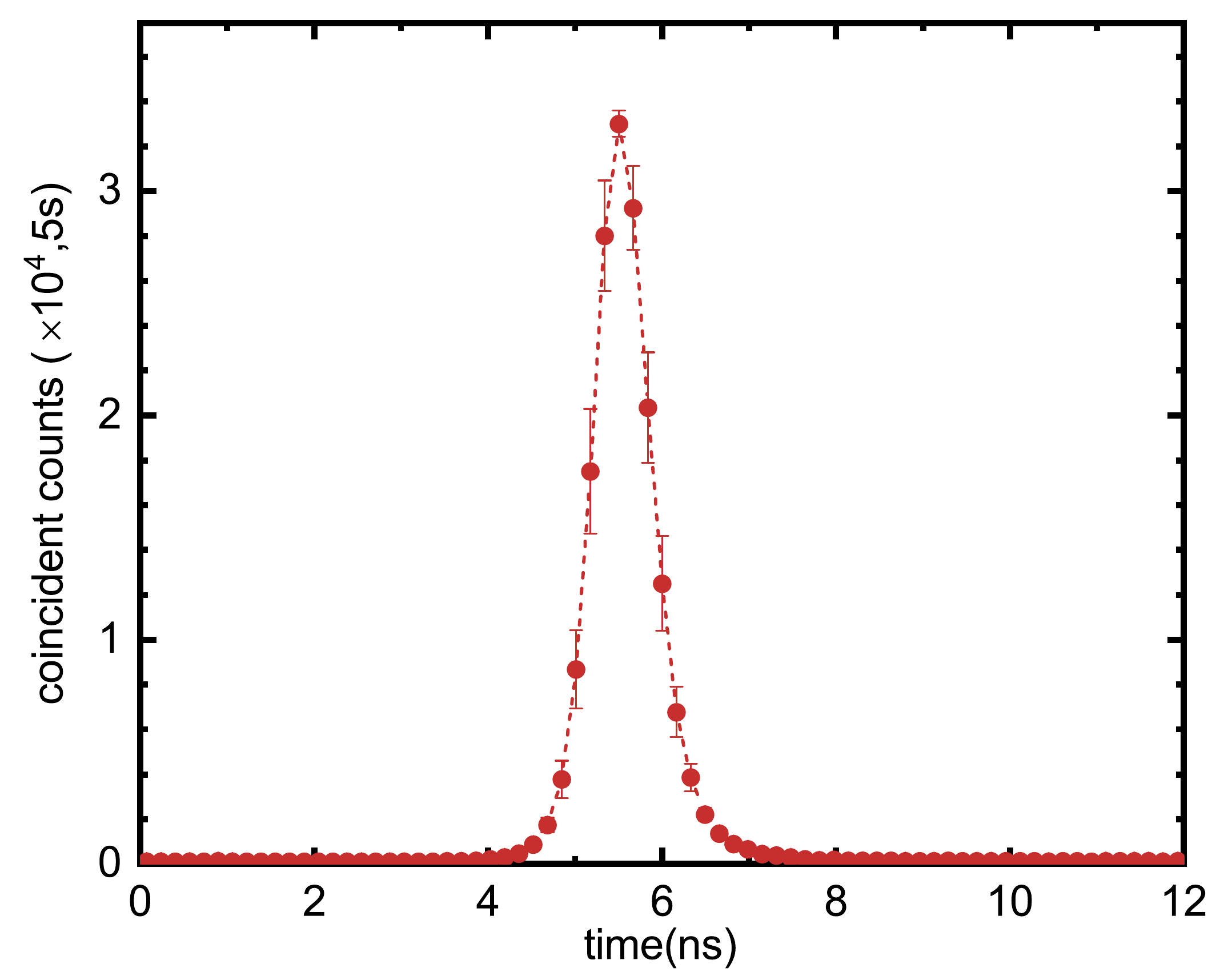}
\caption{\label{coincidence} Coincident counts as a function of the time delay between idler  and signal photons. The dots with error bar represent the mean values and standard deviation of three measured results.}
\end{center}
\end{figure}

Figure \ref{spdc}  is a schematic representation of our single photon source. Single photons are produced by spontaneous parametric down-conversion \cite{Kwiat1995,Burnham1970,Li2021} (SPDC). The laser with wave length of $\lambda_p=397.5~nm$ and power of $60~mW$ is pumped into a periodically poled KTP (PPKTP) crystal. In order to obtain higher spectral brightness of the single photons, we adopted the technique of cavity-enhanced SPDC\cite{Li2021}. A Fabry-P\'erot cavity is formed by the coatings on the end faces of the PPKTP. The front face has a highly reflective (HR) coating with the reflectivity of $99.9\%$ at $795~nm$, while the rear face has a partially reflective (PR) coating with the reflectivity of $85\%$ at $795~nm$. The PPKTP, which is in type-II phase matching, produces a pair of collinear $\lambda=795~nm$ photons with orthogonal polarizations. The paired photons are separated by a PBS, and the reflected (transmitted) photons are denoted as signal (idler) photons.


We use a Hanbury Brown-Twiss (HBT) method \cite{Hanbury2009} to measure the second order correlation function $g^{(2)}_c$ of our single photon source. The idler photons are sent to a single photon detector (Excelitas, SPCM-AQ4C) denoted as D1.  The signal photons, which are delayed by additional length of optical fiber, pass through a 50:50 beamsplitter (BS) and then are detected by two detectors (D2 and D3; Excelitas SPCM-AQ4C). D2 and D3 are connected to a photon correlation module (Becker \& Hickle GmbH, DPC-230). When D1 receives a photon, it triggers a digital delay generator (Standford Research System, DG645). Then the generator sends a square pulse to the correlation module. This module records the starting time of photons measurement. Then the coincidence counts can be measured as a function of the time delay between the signal and idler photons, and the results are shown in Fig. \ref{coincidence}. From this figure, we can notice that the coherence time of the signal photons is clearly shorter than $3\ ns$. So it is safe to set the time $12\ ns$ as the width of the correlation windows in the following calculation of $g^{(2)}_c$.

 The correlation module records the time-tagged data of photons detected by D1, D2 and D3. We denote $N_{1}$ as the count of the photons detected by D1. $N_{12}$ ($N_{13}$) is the coincident count of D1 and D2 (D3), while $N_{123}$ is the coincident count of D1, D2 and D3. Then the second order correlation function $g^{(2)}_c$ can be calculated by \cite{Grangier1986,Bocquillon2009}
\begin{equation}
\begin{split}
\label{g2}
g^{(2)}_c=\frac{N_{1}N_{123}}{N_{12}N_{13}}.
\end{split}
\end{equation}
For an ideal single photon source, $g^{(2)}_c=0$. However, in general, the photon source can be considered as a single photon source if $g^{(2)}_c\in[0,0.5]$.  We derive $g^{(2)}_c=0.094 \pm 0.011$ for our single photon source. In our experiment, the measuring time is set as $5s$, and the result is average over $12$ repeated measurements.

After the estimation of $g^{(2)}_{c}$ of the SPDC, the signal photons are sent to the experimental system of Fig. 2 in the main text. In our experiment of measuring  propagators, each detection of the signal photons on the ICMOS camera is triggered by the idler photon. We connect the ICMOS camera to a digital delay generator. When the camera receives a square pulse from the digital delay generator, it will open a gate for detecting the signal photon. We set the width of the detecting gate to be $12\ ns$, which matches the correlation window in the derivation of $g^{(2)}_{c}$.

Alternatively, the $g^{(2)}_{c}$ can also be determined, in principle,  from measurements using the ICMOS camera. However, it is not practice due to the extensive time required for the experiment.  To measure $g^{(2)}_{c}$ factor with the ICMOS camera, it is necessary to record each gate event of the ICMOS as an individual image with a size of pixels (each with 12-bit greyscale) which results in a file size of 3.32Mb. Due to the large amount of data generated, processing the data becomes challenging. Furthermore, collecting this data requires a significant amount of time since the memory of the camera is limited.
On the other hand, the measurement method we adopted is an efficient and suitable approach for determining the $g^{(2)}_{c}$ factor in Eq.(\ref{g2}). In our experiment, the ICMOS camera exhibits excellent single photon detection performance due to its low readout noise. The duration of each trigger event of the ICMOS camera is 12ns, which is equivalent to the width of the correlation windows used in the calculation of $g^{(2)}_{c}$. Consequently, the photons measured with the ICMOS camera meet the criteria for single photon detection.

\subsection{Phase modulation of the SLM}

The SLM1  can perform a phase shift at a specific position. Because the SLM1 uses liquid crystals for phase modulation, it can shift the light phase along  an axis. In our case, the modulation axis  is in the horizontal direction. We first use an HWP to rotate the light from horizontal polarization $|0\rangle$ to diagonal polarization $|+\rangle=\frac{1}{\sqrt{2}}(|0\rangle+|1\rangle)$.
Then the SLM1 performs a $\pi$ phase shift on $|0\rangle$ ($|H\rangle$) at position $x$. The polarization after this modulation can be written as,
$\frac{1}{\sqrt{2}}(e^{i\pi}|0\rangle+|1\rangle)
=\frac{1}{\sqrt{2}}(-|0\rangle+|1\rangle)
=-|-\rangle,$
 which is equivalent to the operator $U=e^{-i\frac{\pi}{2} \hat{\pi}_{x} \hat{\sigma}_{y}}$. The reflected lights from the SLM1 pass the HWP again. The HWP performs a reverse transformation that rotates $|-\rangle$ to $|1\rangle$ and $|+\rangle$ to $|0\rangle$.

The phase modulation of the SLM1 may lead to a fluctuation described by  $e^{i\epsilon\hat{I} \hat{\sigma}_{y}}$ with $\epsilon$ being a small angle, which corresponds to the background grey level displayed on the SLM1.   To eliminate this error,  we have measured this background and deducted it from the measurement of
$\langle f|\hat{\sigma_{x}}|f\rangle$ and $\langle f|\hat{\sigma_{y}}|f\rangle$.

\subsection{The global phase factor of the propagators}

    Shifting an arbitrary global phase to the propagator $K$ will not change the physical observations. Therefore, to clearly compare the theoretical and observed propagators in Figs.~\ref{Propagator data} and~\ref{Classical path}, we have shifted  a global phase  $e^{i(\beta_{m}-\beta_{t})}$ on the theoretical propagator. Here the phase angles $\beta_{m}=\arctan{(\frac{Re[K_{m}]}{Im[K_{m}]})}$ and $\beta_{t}=\arctan{(\frac{Re[K_{t}])}{Im[K_{t}]})}$, where the measured (theoretical) propagator is denoted as $K_{m}$ ($K_{t}$).

\subsection{Calculations of classical trajectories with the PLA}

We  here show that the wave functions do not have to be measured  to calculate classical trajectories with experimental data.  From Eq.(\ref{Kxx}), we have $K'(x_b,t_b;x,t)=\psi^{*}(x_b,t_b)\psi(x,t)K(x_b,t_b;x,t)$, where $K(x_b,t_b;x,t)=\mathcal{N}_b\exp(\frac{i}{\hbar}S_b(x))$;  and $K'(x,t;x_a,t_a)=\psi^{*}(x,t)\psi(x_a,t_a)K(x,t;x_a,t_a)$, where $K(x,t;x_a,t_a)=\mathcal{N}_b\exp(\frac{i}{\hbar}S_a(x))$.  $\mathcal{N}_a$ and $\mathcal{N}_b$ are normalization constants and are independent of $x$ [e.g, Eq.(\ref{Kf}) and Eq. (\ref{Kh})]. Therefor, we obtain

\begin{equation}
\Pi(x,t)=\frac{\Pi^\prime (x,t)}{\psi^{*}(x_{b},t_{b})\psi(x_{a},t_{a})|\psi(x,t)|^{2}}=\frac{\mathcal{M}^\prime (x,t)}{F(\gamma)},
\end{equation}
where $\mathcal{M}^\prime (x,t)={\Pi^\prime (x,t)}/{|\Pi^\prime (x,t)|}$ with $\Pi^\prime (x,t)=K'\left(x_{b}, t_{b}; x, t\right)K'\left(x, t; x_{a}, t_{a}\right)$.
Here $F(\gamma)=\mathrm{e}^{i\gamma}/\mathcal{N}^2$ with
$\mathrm{e}^{i\gamma}=\frac{\psi(x_{a},t_{a})\psi^{*}(x_{b},t_{b})}{|\psi(x_{a},t_{a})\psi^{*}(x_{b},t_{b})|}$ being an $x$ independent function.
Similarly, from Eq. (\ref{Kxx2}), we may obtain $\Pi(x,t)=\mathcal{M}^{\prime\prime} (x,t)/F(\gamma^{\prime\prime})$, where
$\mathcal{M}^{\prime\prime} (x,t)= \Pi^{\prime\prime} (x,t)/|\Pi^{\prime\prime} (x,t)|$ with $\Pi^{\prime\prime} (x,t)=K^{\prime\prime}\left(x_{b}, t_{b}; x, t\right)K^{\prime\prime}\left(x, t ; x_{a}, t_{a}\right)$ and $\mathrm{e}^{i\gamma^{\prime\prime}}=\frac{\psi(x_{a},t_{a})\psi^{*}(x_{b},t_{a})}{|\psi(x_{a},t_{a})\psi^{*}(x_{b},t_{a})|}$.  Therefor, although the measurement of the wave function is required for the detection of propagators, it is not needed in calculating the extremum of the $\Pi(x,t)$.
Derived from Eq.(\ref{PLA}), the PLA  can be expressed as
\begin{equation}
\label{PLA_E}
\frac{\partial}{\partial x} Re[\Pi(x,t)]=0,\ \ \ \frac{\partial}{\partial x} Im[\Pi(x,t)]=0.
\end{equation}
Eq.(\ref{PLA_M}) in the main text can be derived from Eq.(\ref{PLA_E}).

\begin{figure*}
\begin{center}
\includegraphics[width=14cm]{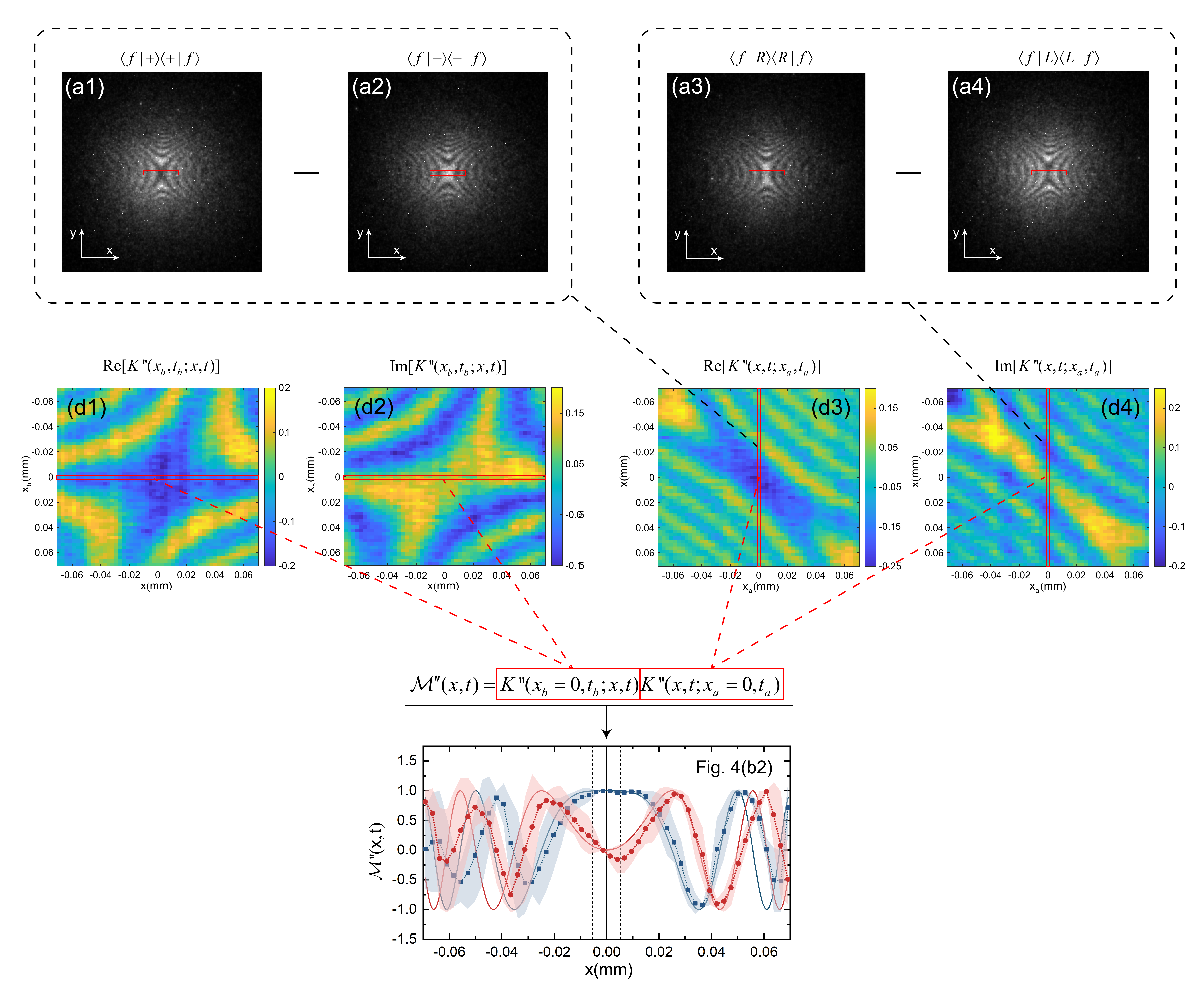}
\caption{\label{raw_data} The calculations from raw data to $\mathcal{M}''(x,t)$.}
\end{center}
\end{figure*}

\subsection{Numerical differentiation method}

In our experiment, the PLA  is expressed as Eq. (5).
We use the Richardson's extrapolation as the numerical differentiation to solve Eq.(5).  The Richardson's extrapolation  can be expressed as
\begin{equation}
\begin{split}
\frac{\partial f(x)}{\partial x}=\frac{-f(x+2 h)+8 f(x+h)-8 f(x-h)+f(x-2 h)}{12 h},
\end{split}
\end{equation}
where $h$ is the step size. By using this numerical method, we get the derivative of the contribution of the path from the measured results.
The read out noise of the ICMOS camera may affects the precision of the differential algorithm in data analysis, so
we have adopted the calculation of moving average to reduce the noise, which averages the nearest points of $\mathcal{M}''(x,t)$ on $x$ axes.
Then, we select the $x$ position at which $|\frac{\partial\mathcal{M}^{\prime\prime}(x)}{\partial x}|^{2}$ gets the minimum value. In other words, we transform the condition $|\frac{\partial\mathcal{M}^{\prime\prime}(x)}{\partial x}|^{2}=0$ into an approximation condition $\operatorname{Min}[|\frac{\partial\mathcal{M}^{\prime\prime}(x)}{\partial x}|^{2}]$.\\
\\

\subsection{The calculations from raw data to $\mathcal{M}''(x,t)$}

In this section, we demonstrate the calculations involved in obtaining $\mathcal{M}''(x,t)$ from the raw images captured by the ICMOS camera, taking the results in Fig. 4(b2) as an example. The schematic diagram of the calculations is shown in Fig. \ref{raw_data}. To obtain $K''(x,t;x_a,t_a)$, we project single photons to four polarization bases, $\{|+\rangle,|-\rangle,|R\rangle,|L\rangle\}$, and detect them using the ICMOS camera. The raw images captured by the camera on these four bases are shown in (a1)-(a4) of Fig. \ref{raw_data}, with an exposure time of $400~s$. We extract the intensity (12-bit gray scale) at the area of $\{-1*pixel\le y\le 1*pixel,-26*pixels\le x\le 26*pixels\}$, shown as red frames in Figure \ref{raw_data} (a1)-(a4)) in the images.
 We then obtain $Re[K''(x,t;x_a,t_a)]$ ($Im[K''(x,t;x_a,t_a)]$) by taking the difference between the images of $\langle f|+\rangle\langle +|f\rangle$ and $\langle f|-\rangle\langle -|f\rangle$ ($\langle f|R\rangle\langle R|f\rangle$ and $\langle f|L\rangle\langle L|f\rangle$). The matrixes of $Re[K''(x,t;x_a,t_a)]$ and $Im[K''(x,t;x_a,t_a)]$ are shown in Fig. \ref{raw_data} (d3) and (d4), respectively.
 To estimate the classical trajectory from $x_a=0$ to $x_b=0$, we extract $K''(x_b=0,t_b;x,t)$ and $K''(x,t;x_a=0,t)$ (shown as red frames in Figure \ref{raw_data} (d1)-(d4)) from the matrixes of $K''(x_b,t_b;x,t)$ and $K''(x,t;x_a,t)$. Finally, we derive $\mathcal{M}''(x,t)$  from the product of $K''(x_b=0,t_b;x,t)$ and $K''(x,t;x_a=0,t)$.

\subsection{GRIN lens}\label{grin_met}

In our experiment, a gradient refractive index (GRIN) lens is used to form a harmonic potential for photons.
The refractive index of the GRIN lens  can be expressed as
\begin{equation}
\begin{split}
n(r)=n_0(1-\frac{1}{2}A r^2),
\end{split}
\end{equation}
where $n_0=1.643$ is the maximum refractive index of the sample, $r=\sqrt{x^2+y^2}$, and $A=0.043 ~mm^{-2}$ is the gradient constant.  The effective potential for the photons is a  harmonic potential $V(r)=\frac{1}{2}m\omega^{2}r^2$, where  $\omega={2\pi}/{T}$ with $T=30.26~mm/c$ being the length of one cycle of the light travel in the GRIN lens. With the constraint on one dimension $y=0$, we have $V(x)=\frac{1}{2}m\omega^{2}x^2$.

\begin{figure}
\begin{center}
\includegraphics[width=8cm]{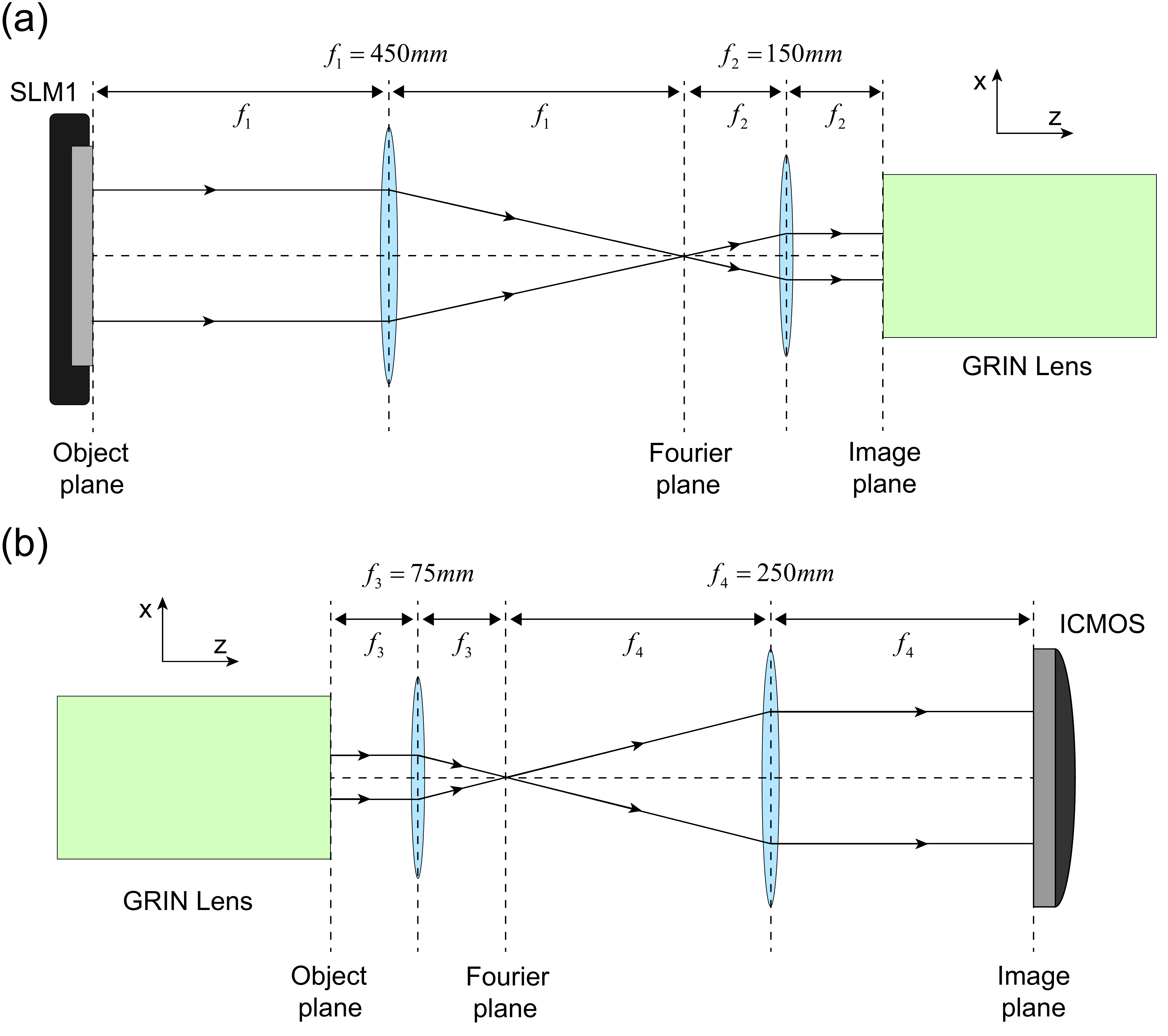}
\caption{\textbf{\label{Imaging} Imaging systems.} (a) Imaging system for the SLM1. (b) Imaging system for the ICMOS camera.    }
\end{center}
\end{figure}
\subsection{Imaging systems}

Our imaging system for the SLM1, consisting of two lenses, is used to project the wave front to the position of $t_{a}$. The pixel size of the SLM1 is $8~\mu m$. In order to have a higher resolution in our measurement, it is appropriate to have a smaller width of a measuring point. We use two convex lenses ($f_{1}=450~mm$ and $f_{2}=150~mm$) in a $4f$ arrangement, which is shown in Fig. \ref{Imaging}a. The state  is prepared after modulation by the SLM1 at $t_{a}$. We place the surface of the SLM1 at the object plane of the imaging system. Passing through this two lenses, the state will be projected at the imaging plane, which is the incident surface of the GRIN lens. By using this imaging system with a magnification $1/3$, the width of the measuring point is decreased to $2.67~\mu m$.

The pixel size of our ICMOS camera is $9~\mu m$. In order to match the measuring width of $x_{a}$ and $x_{b}$ in the same size as much as possible, we also use a $4f$ imaging system (Fig. \ref{Imaging}b) with a magnification  $10/3$ to image the wave front at $t_{b}$ to the ICMOS camera. In this case,  the measuring width of $x_{a}$ ($8.89\ \mu m$) is nearly equal to a pixel of the camera ($9~\mu m$).

\subsection{Robustness of the classical trajectories}

\begin{figure} [h]
\begin{center}
\includegraphics[width=8cm]{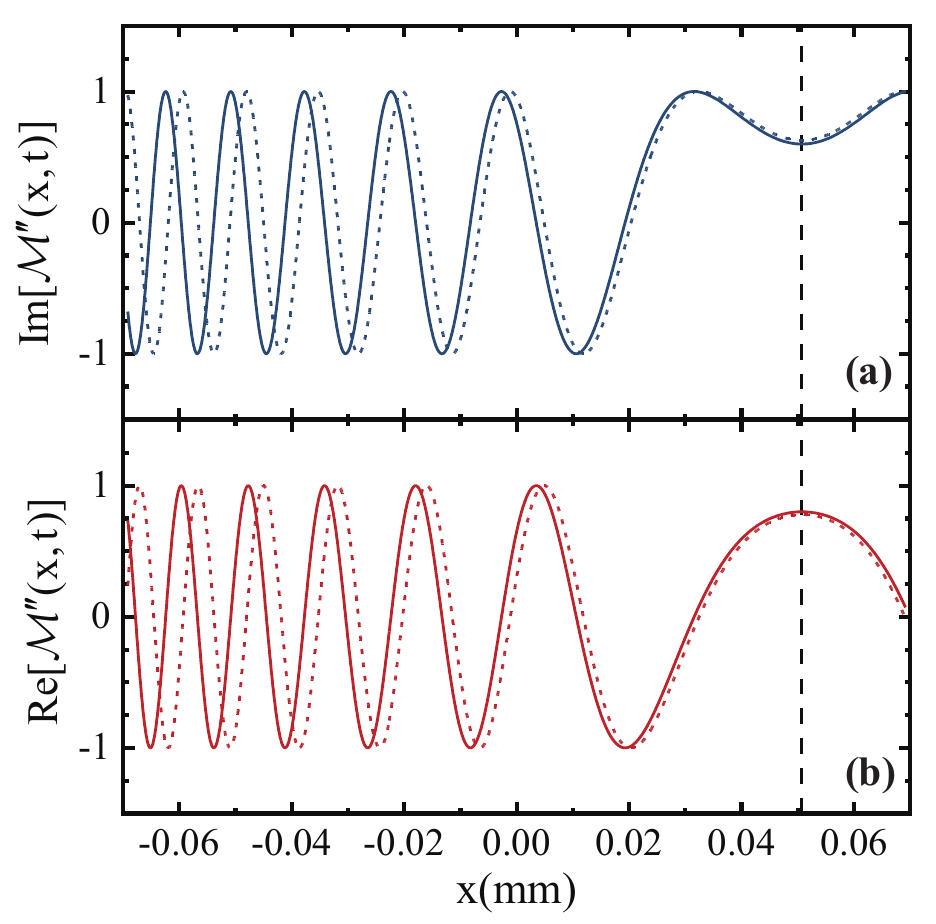}
\caption{\label{t_error} Theoretical comparison of $\mathcal{M}^{\prime\prime}(x,t)$ and $\mathcal{M}^{\prime\prime}_{\epsilon}(x,t-\epsilon)$ in the harmonic potential, where $t=1.1~\pi/\omega$ and $\epsilon=0.003(t_b-t_a)$. \textbf{(a)} and \textbf{(b)} show  the imaginary and real components of $\mathcal{M}^{\prime\prime}(x,t)$ (solid) and $\mathcal{M}^{\prime\prime}_{\epsilon}(x,t)$ (dashed).
The dashed vertical straight line shows the classical position determined by $\mathcal{M}^{\prime\prime}(x,t)$.}
\end{center}
\end{figure}

 In our experiment, the fluctuations of some experimental parameters, such as the control time and the relative distance between optical elements, can  affect the measurement of the propagators, but the classical trajectories calculated with the PLA is very robust.  Here we take the uncertainty of control time as an example.  We calculate the ideal $\mathcal{M}^{\prime\prime}(x,t)$ and a deviation of the propagator   $\mathcal{M}^{\prime\prime}_{\epsilon}(x,t)=\frac{K^{\prime\prime}(x_b,t_b;x,t)K^{\prime\prime}(x,t-\epsilon;x_a,t_a)}{|K^{\prime\prime}(x_b,t_b;x,t)K^{\prime\prime}(x,t-\epsilon;x_a,t_a)|}$
 induced by a  small deviation of time $\epsilon=0.003(t_b-t_a)$. The comparison is plotted in Fig. \ref{t_error}.  It shows that the deviation of $\mathcal{M}^{\prime\prime}(x,t)$ for a very small change of the intermediate time $t$ is small near the classical position but large in regions far away from the classical position. The fidelity of the measured propagators is given by $F=|\int^{L/2}_{-L/2} [\mathcal{M}^{\prime\prime}(x,t)]^{*}\mathcal{M}^{\prime\prime}_{\epsilon}(x,t) dx|^2=0.6853$, where $L$ is the measuring range of $x$. The difference between the classical positions calculated from $\mathcal{M}^{\prime\prime}(x,t)$ and $\mathcal{M}^{\prime\prime}_{\epsilon}(x,t)$ is smaller than $10^{-3}L$. Thus, the classical trajectories calculated with the PLA based on experimental data  are very robust against perturbations, while the fidelity of the measured $\mathcal{M}^{\prime\prime}(x,t)$ is strongly affected.

\subsection{The measuring errors}

In our experiment, the control of experimental parameters is not perfect and may cause errors to the results of the propagators. Here we briefly address these errors.

The errors from the 4f-systems. The 4f-systems are one of the crucial equipments of our experiment, which determine the fidelity of transiting the assigned wave front of the photons. The error of the distance between two lens of the 4f-system would cause a longitudinal shifting and a transverse magnification to the image plane. Besides, the error of the longitudinal distance between the 4f-systems and the SLM, the GRIN lens and the ICMOS camera, would affect the control of the initial and terminal time of the measured propagators. These errors impact the precision of the propagator in the region far away from the classical position more obviously.

The error of the length of the GRIN lens. The length of the GRIN lens represents the evolving duration of the propagators in the harmonic potential. This error affects the precision of the control time. As we have discussed in Methods, even a small change of this length would cause a great decrease of the fidelity of the measured $\mathcal{M}''(x,t)$.

\bigskip
\textbf{Acknowledgements:} We thank Z. Y. Zhou for helpful discussions. This work was supported by  the Key-Area Research and Development Program of GuangDong Province (Grants No. 2019B030330001 (S.-L.Z. and H.Y.) and No.2020A1515110848 (S.Z.)), the National Key Research and Development Program of China (Grants No.2020YFA0309500 (S.Z.) and No.2022YFA1405300 (S.-L.Z.)),  the National Natural Science Foundation of China (Grants No.12225405(H.Y.), No.12004120 (Y.W.), No.62005082 (J.L.),  No.12074180(S.-L.Z.), and No.U20A2074 (H.Y.)), and the Innovation Program for Quantum Science and Technology (Grant No. 2021ZD0301705 (H.Y. and S.-L.Z.)).

\textbf{Author contributions:}
  Y.-L.W. and S.-L.Z. developed the theory, Y.-L.W, Y. W., S.Z., H.Y., and S.-L.Z. designed the experiment. Y.-L. W, Y.W., L.-M.T., S.Z., J.L., and J.-S. D.carried out the experiments. Y.-L.W., Y.W., H.Y., and S.-L.Z. conducted the raw data analysis. Y.-L.W. H.Y., and S.-L.Z. wrote the paper, and all authors
discussed the content of the paper. H.Y. and S.-L.Z. supervised the project.

\noindent
\textbf{Competing interests:} The authors declare no competing financial interests.



\end{document}